\documentclass[11pt]{article}

\usepackage{epsfig,subfigure}
\usepackage{amssymb, amsmath}
\usepackage{color}
\def\QED{\mbox{\rule[0pt]{1.5ex}{1.5ex}}}

\usepackage{graphicx}

\usepackage{amssymb}
\usepackage{amsmath,amsfonts,amssymb}
\usepackage{verbatim}
\usepackage{stfloats}
\usepackage[bookmarks=false]{}

\newtheorem{theorem}{Theorem}

\begin{document}
\date{}
\title{Index Coding Capacity: How far can one go\\ with only Shannon Inequalities?}
\author{ \normalsize Hua Sun and Syed A. Jafar \\
{\small University of California Irvine, Irvine, CA 92697}\\
{\small \it Email: \{huas2, syed\}@uci.edu}
}
%Solving Index Coding Problems \\with Alignment and Conflict Graphs
\maketitle

\begin{abstract}
An interference alignment perspective is used to identify the simplest instances (minimum possible number of edges in the alignment graph, no more than 2 interfering messages at any destination) of index coding problems where non-Shannon information inequalities are necessary for capacity characterization. In particular, this includes the first  known example of a multiple unicast (one destination per message) index coding problem where non-Shannon information inequalities are shown to be necessary. The simplest multiple unicast example has 7 edges in the alignment graph and 11 messages. The simplest multiple groupcast (multiple destinations per message) example has 6 edges in the alignment graph, 6 messages, and 10 receivers. For both the simplest multiple unicast and multiple groupcast instances,  the best outer bound  based on only Shannon inequalities is $\frac{2}{5}$, which is tightened to $\frac{11}{28}$ by the use of the Zhang-Yeung non-Shannon type information inequality, and the linear capacity  is shown to be $\frac{5}{13}$ using the Ingleton inequality.  Conversely, identifying the minimal challenging aspects of the index coding problem  allows an expansion of the class of solved index coding problems up to (but not including) these instances.

\end{abstract}
\newpage

\section{Introduction}

The capacity of a general ``Index Coding" communication network is one of the most intriguing problems in network information theory. The index coding problem is  simple to describe (only one link of finite capacity) but  difficult to solve (remains open in general), it is the original setting for interference alignment \cite{Birk_Kol, Jafar_FnT} but is only starting to be explored from an interference alignment perspective \cite{Jafar_TIM, Maleki_Cadambe_Jafar}, and while it is a network coding problem itself, it has been shown to be  representative of \emph{all} (including non-linear) network coding instances \cite{Rouayheb_Sprintson_Georghiades, Effros_Rouayheb_Langberg}. The index coding problem has also been shown recently to be essentially equivalent (up to linear solutions) to the so called  topological interference management problem \cite{Jafar_TIM}, where the degrees of freedom of a partially connected wireless interference network or the capacity of a partially connected wired network are investigated with only a knowledge of the network topology available to the transmitters. As such the index coding problem presents an opportunity to tackle some of the  fundamental challenges lying at the intersection of several open problems in network information theory.

\subsection{Prior Work}
Since its introduction in 1998 \cite{Birk_Kol}, many interesting instances of the index coding problem have been studied  from coding theoretic,  graph theoretic, and information theoretic  perspectives, leading primarily to a variety of inner bounds (achievable schemes). The earliest inner bound, obtained  by Birk et al. in \cite{Birk_Kol,Yossef_Birk_Jayram_Kol_Trans}, is the   clique cover of an index coding side information graph. The clique cover and its standard LP generalization, the fractional clique cover, correspond to orthogonal scheduling (analogous to TDMA/FDMA) in the parlance of interference networks \cite{Jafar_TIM} ---  only non-interfering groups of users are simultaneously scheduled for transmission. A  linear programming inner bound is introduced by Blasiak et al. in \cite{Blasiak_Kleinberg_Lubetzky_2010},  based on higher order sub-modularity and coincides with fractional hyperclique-cover number (reduces to fractional clique-cover number for multiple unicast instances). The fractional clique cover inner bound is generalized to a partition multicast inner bound by Tehrani et al. in  \cite{Tehrani_Dimakis_Neely}. The partition multicast approach corresponds to CDMA in interference networks --- pseudo-random precoding sequences are used with the length of the sequences chosen to be just enough to provide each receiver enough equations so it can resolve all symbols from the transmissions that it can hear concurrently with its desired transmission \cite{Jafar_TIM}.  The local chromatic number is proposed as an inner bound by Shanmugam et al. in  \cite{Alex_local} based on viewing the index coding problem as a vector assignment problem. This is equivalent to a restricted form of interference alignment, sometimes known as \emph{one-to-one} alignment (as opposed to the more general concept of \emph{subspace} alignment). For index coding problems that correspond to undirected graphs (equivalently, bidirected graphs) all of the inner bounds mentioned above --- the fractional clique cover, partition multicast, the linear programming bound based on higher-order submodularity, and the local chromatic number ---  are equivalent. They are also generally suboptimal.  A family of  undirected graph based index coding problems with $n$ nodes is presented by Blasiak et al. in \cite{Blasiak_Kleinberg_Lubetzky_2010} where the index coding capacity is 1/3,  but the best rate achievable by all  inner bounds mentioned above is no more than $O\left(\frac{1}{n^{1/4}}\right)$. The inner bound that is tight for this class of index coding problems is  given by the min-rank function, which was originally introduced by Yossef et al. in \cite{Yossef_Birk_Jayram_Kol_Trans} and corresponds to the optimal scalar linear solution to the index coding problem. However, even the min-rank bound is  known to be suboptimal. First, it corresponds to scalar linear coding capacity, which is generally outperformed by vector linear coding \cite{Alon_Hasidim_Lubetzky_Stav_Weinstein, Rouayheb_Sprintson_Georghiades}. Second, while the min-rank bound can be extended as shown in \cite{Lubetzky_Nonlinear, Jafar_TIM} to find the best  \emph{vector} linear coding solution for a given number of symbols per message,   it has also been shown by Blasiak et al. and Rouayheb et al. for multiple groupcast index coding problems \cite{Blasiak_Kleinberg_Lubetzky_2011, Rouayheb_Sprintson_Georghiades}, and by Maleki et al. for multiple unicast index coding problems \cite{Maleki_Cadambe_Jafar}, that the best linear schemes are still not optimal in general because they can be outperformed by non-linear schemes for certain instances of the index coding problem.  Therefore, none of the inner bounds discussed so far has the potential to be generally optimal for the index coding problem.

\subsection{Recent Progress: Random Coding and Interference Alignment Approaches}
Two recently proposed approaches, while still in their infancy, offer new hope by bringing in new machinery to attack the index coding problem. These are the random coding approach by Arbabjolfaei et al. in \cite{Arbabjolfaei_region}, and the interference alignment perspective of Jafar et al. in \cite{Jafar_TIM, Maleki_Cadambe_Jafar}. Random coding  has been a universal ingredient of capacity optimal schemes, and therefore presents a potentially powerful ``hammer"  to the ``nail" of index coding \cite{Arbabjolfaei_region}. The random coding approach of \cite{Arbabjolfaei_region} has so far settled the capacity \emph{region} for all instances of the multiple unicast index coding problem with 5 or fewer messages.\footnote{The capacity of an interesting sub-class of multiple unicast index coding problems with 6 messages, corresponding to the topologies that can arise in a 6-cell network, is settled by Naderializadeh et al. in \cite{Naderi_Avestimehr}.} The interference alignment perspective presents a different kind of hammer (or perhaps a ``jack knife" since it takes a finer view of the problem \cite{Arbabjolfaei_region}),  that has been extremely successful in wireless networks where  a variety of interference alignment schemes, both linear and non-linear, have been developed to obtain degrees of freedom characterizations \cite{Jafar_FnT}. In addition to symmetric instances of the index coding problem such as neighboring antidotes, neighboring interferers, $X$ networks, and cellular topologies motivated by the topological interference management problem \cite{Maleki_Cadambe_Jafar}, the interference alignment perspective has so far settled the symmetric capacity of the class of multiple \emph{groupcast} (which includes unicast as a special case) index coding problems where each alignment set has either no cycles or no forks \cite{Jafar_TIM}.

Admittedly, the accomplishments of either approach thus far are infinitesimal relative to the full scope of the general index coding problem. However, the full strength of neither the random coding approach nor the interference alignment approach has yet been exhausted. Indeed it is apparent that both approaches have plenty of room to expand through capacity characterizations of increasingly broader classes of the index coding problem. Continued efforts towards such expansions are therefore well motivated. At the same time, it is also important to understand the limitations of these approaches  by identifying the challenges that lie ahead.
To this end, we note that all index coding capacity results obtained so far from random coding and interference alignment perspectives, have relied on only Shannon inequalities for the outer bounds. One indication of a substantial challenge could be the necessity of non-Shannon inequalities. Remarkably, for multiple unicast index coding problems it is not known whether non-polymatroidal (non-Shannon, Ingleton) inequalities are ever necessary. Therefore, we would like to find out if  instances of multiple unicast index coding exist where Shannon inequalities do not suffice, and if so, then we would like to identify the simplest possible such instance. The emphasis on simplicity is important for the challenging aspects to be as broadly relevant as possible.

\subsection{Non-Shannon Inequalities}
 To characterize the information-theoretic/linear capacity of  communication networks, it is important to understand the fundamental limitations of  entropy/vector spaces,  in the form of information inequalities/linear rank inequalities. Since linear coding schemes are only a  special case of all possible coding schemes,  linear rank inequalities are a proper subset of information inequalities. It is well known that both entropy space and vector space satisfy the basic polymatroidal axioms, which is equivalent to the non-negativeness of Shannon information measurements, also known as Shannon-type information inequalities or basic inequalities \cite{Y_ITNC}. With up to 3 random variables/subspaces, all the information inequalities/linear rank inequalities coincide with the polymatroidal axioms \cite{HRSK_Inequ}. However, when the number of elements increases to 4, both information inequalities and linear rank inequalities involve additional constraints beyond the polymatroidal axioms. For information inequalities, the first non-Shannon-type information inequality with 4 random variables was discovered in 1998 by Zhang and Yeung \cite{ZY_Nonshannon}, followed by many others, e.g., \cite{Makarychev}\cite{Matus_Infty}\cite{DFZ_Nonshannon4}. It is shown by Matus that even with only 4 random variables, the list of non-Shannon-type information inequalities is infinite \cite{Matus_Infty}. For linear rank inequalities, Ingleton \cite{Ingleton} showed a set of inequalities that is not implied by basic inequalities, but when combined with them, constitutes  a complete characterization of all linear rank inequalities that must be satisfied by 4 subspaces of a vector space \cite{HRSK_Inequ}. However, Ingleton inequalities are still not enough to go beyond 4 subspaces \cite{Kinser}\cite{DFZ_Rank}. Recent work finds all the linear rank inequalities for 5 subspaces (there are 24 such inequalities in addition to the Shannon and Ingleton inequalities), while cases with more than 5 subspaces are still open \cite{DFZ_Rank}.

 All these mathematical inequalities have found their usage and correspondence in network coding studies \cite{DFZ_Matroids}\cite{CG_Duality}.  Dougherty et al. in \cite{DFZ_Nonshannon} have shown that Shannon inequalities are not always sufficient for the general network coding problem. For index coding problems the insufficiency of Shannon inequalities is established by Blasiak et al. in \cite{Blasiak_Kleinberg_Lubetzky_2010}, albeit only in the context of multiple \emph{groupcast} index coding where each message is desired by multiple receivers. For multiple unicast index coding, however, no example is known that shows that Shannon inequalities are insufficient. For instance, the general outer bound for multiple unicast index coding presented in Theorem 1 of \cite{Arbabjolfaei_region}  is based directly on Shannon inequalities and it is noted afterwards that it is not known whether the outer bound is tight in general. It is also notable that this bound is found to be tight for all instances of multiple unicast index coding with 5 or fewer messages.
 For further details about information inequalities and linear rank inequalities, we refer to the excellent tutorials in \cite{Y_Entropy}\cite{C_Entropy} and references therein.

Since the notation and definitions used in this work are the same as in \cite{Jafar_TIM}, we proceed directly to the results. The relevant definitions from \cite{Jafar_TIM} are summarized for the sake of completeness in Appendix  \ref{sec:def}.

\section{Results}
\subsection{Criteria for the Simplest Example}

Our first goal is to prove that Shannon inequalities are insufficient even for multiple unicast index coding, so that the outer bound in Theorem 1 of \cite{Arbabjolfaei_region} cannot be tight in general. For such a result, the simplest example is the most powerful. Therefore, we would like to find an example  that involves only those features that would make it a part of any interesting class of index coding problems. For instance, a multiple unicast example would prove the insufficiency result for both multiple unicast and the multiple groupcast settings, because multiple unicast settings are contained within the class of multiple groupcast settings. So our example must be a multiple unicast setting. This is especially critical because the insufficiency of Shannon inequalities is already shown for the more general groupcast setting \cite{Blasiak_Kleinberg_Lubetzky_2010}.

Continuing  the  thought, even within the class of multiple unicast index coding problems, we would like to identify the simplest example possible, for the result to hold as broadly as possible. The idea of  `simplicity' can be quite subjective. However, coming from an interference alignment perspective, we find it natural to interpret it in terms of the `type' and `number' of  edges in the alignment graph, as defined in \cite{Jafar_TIM}.

The type of an alignment edge refers to whether the alignments that it \emph{demands}\footnote{It is important to distinguish between an alignment \emph{demand}, which represents an edge in the alignment graph, and an alignment \emph{solution} which could be the optimal vector space assignment. For example,  even if all interference-alignment demands are one-to-one, the optimal solution may require subspace alignments strictly beyond one-to-one alignments. Such examples are not uncommon, e.g., one  appears in Fig. \ref{fig:gach} of this paper.} are one-to-one alignments or subspace-alignments. Let us elaborate on this distinction. Consider an interference network where a receiver experiences interference from only two undesired messages. The principle of interference alignment dictates that these two undesired transmissions should try to collectively occupy as small a signal space as possible. Since there are only two interferers, the only way to consolidate the interference space is for them to align with each other as much as possible. So what is demanded is a \emph{one-to-one} alignment. Now, consider a different scenario where the receiver sees interference from three or more interferers. The principle of interference alignment again dictates that these three or more undesired transmissions should try to collectively occupy as small a signal space as possible. Note however, that a direct alignment of any interferer with any other interferer is no longer the only way to consolidate the signals. For instance, one interferer may align itself in the space spanned jointly by the others without even partially aligning with any of them on a one-to-one basis. What is demanded here is the more general notion of \emph{subspace-alignment}. Since subspace-alignment includes one-to-one alignment as a special case, we naturally require that the simplest example should have only one-to-one alignment demands, i.e., no receiver should see more than two interferers.

The number of alignment conditions simply refers to the number of edges in the alignment graph. Since each edge represents a desired alignment, the number of edges roughly corresponds to the number of dependencies among the variables involved. By this understanding, the simplest example is the one with  the minimum number of dependencies, i.e., fewest  edges in the alignment graph. At this point we have identified the  criteria that the simplest example should satisfy.

\bigskip
\noindent{\bf Criteria for the Simplest Example}
\begin{enumerate}
\item Requires non-Shannon inequalities.
\item Is a multiple unicast index coding problem.
\item Each receiver sees no more than 2 interferers (one-to-one alignment demands).
\item Has the minimum number of edges in its alignment graph (among all examples that satisfy the first three criteria).
\end{enumerate}

{\it Remark: }While this definition of simplicity is motivated by the interference alignment perspective, other definitions may be interesting from other perspectives. For example, since the approach taken by \cite{Arbabjolfaei_region} involves solving all multiple unicast index coding problems up to a given number of users, the  simplest example in that sense might be the one with the minimum number of users, disregarding the number of edges in the alignment graph and the restriction to one-to-one alignments.

\bigskip
Note that  it was not known \emph{a priori}  that an example satisfying our  criteria even exists. However, identifying these criteria helps us search for such an example. Our initial motivation in performing this search was to either  find the simplest such example, or to settle the capacity for this entire class of index coding problems. What is remarkable is that in all instances that belong to this class, for which the capacity was previously known, the capacity coincided with the internal conflict bound (Corollary 4.13 of \cite{Jafar_TIM}), which is based only on Shannon inequalities. Specifically, in all such cases that were previously solved, the symmetric capacity was equal to $\frac{\Delta}{2\Delta+1}$, where $\Delta$ is the minimum internal conflict distance of the alignment graph \cite{Jafar_TIM}.  Based on this observation, our initial expectation had been that the conflict bound may be always tight for this class of problems, and therefore Shannon inequalities will be sufficient. However,  the result turns out to be somewhat unexpected. The conflict bound
is indeed found to be tight for all networks in this class \emph{as long as there are no overlapping cycles in the alignment graph}, i.e., cycles that share an edge. However, once we involve overlapping cycles we almost immediately run into the simplest example where non-Shannon inequalities are necessary.

\subsection{The Simplest Example where Non-Shannon Inequalities are Necessary}
\subsubsection{Construction}
\begin{figure}[h]
\begin{center}
\includegraphics[width=6.5 in]{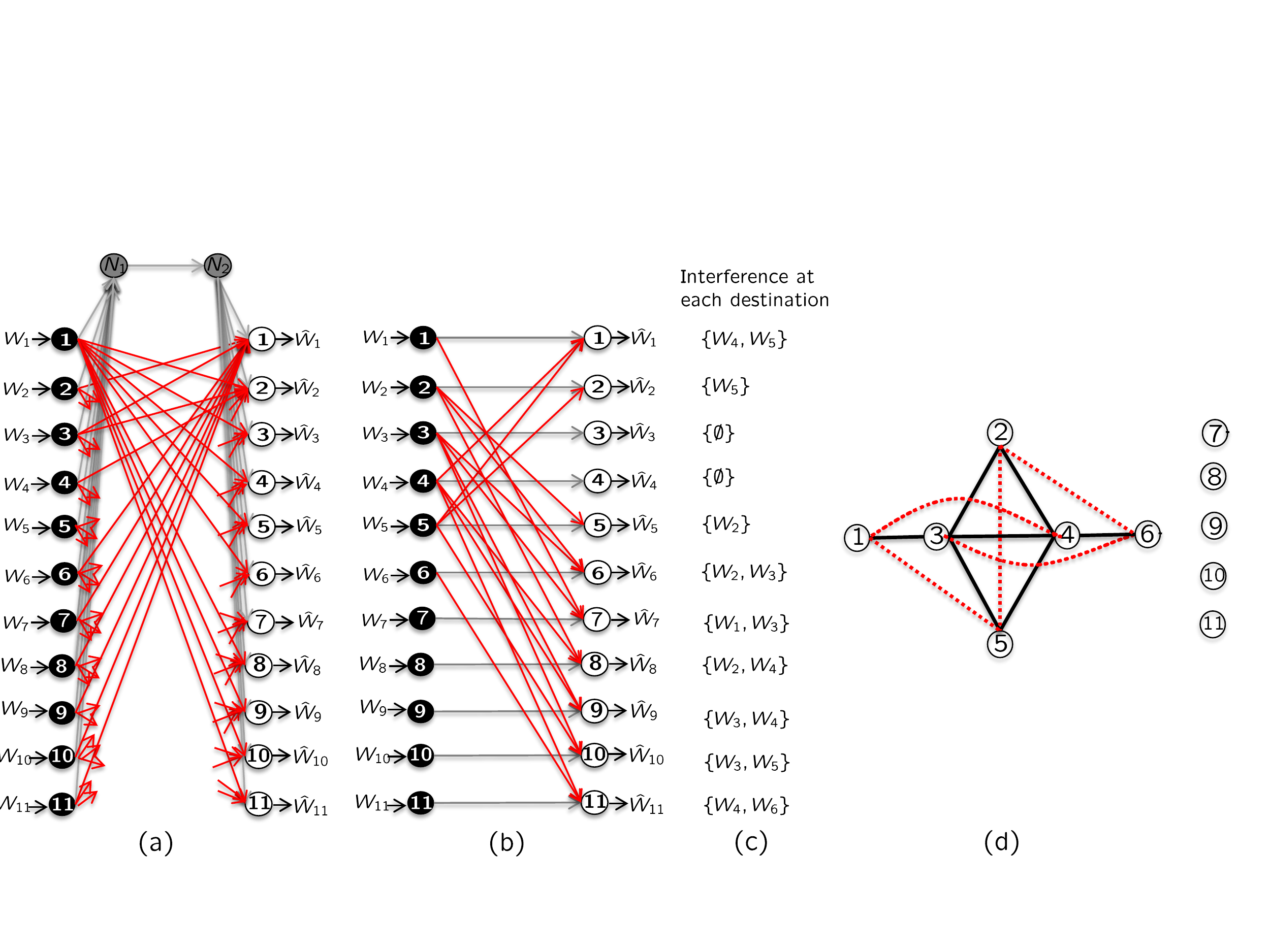}
\caption{ \it\small The Simplest Example of a Multiple Unicast Index Coding Problem where Non-Shannon Inequalities are Necessary.  (a) Red links represent antidotes (not all are shown), and  (b) Red links show all the interference links (complements of the antidote links), and (c) Interfering messages (missing antidotes) at each destination are listed, and (d) Alignment graph (solid black edges) and conflict graph (dashed red edges). Only internal conflicts are shown. }\label{fig:full}
\end{center}
%\vspace{-0.3cm}
\end{figure}

%\begin{figure}[h]
%\begin{center}
%\includegraphics[width=5 in]{topIAic}
%\caption{ \it The Simplest Example of a Multiple Unicast Index Coding Problem where Non-Shannon Inequalities are Necessary.  (a) Red links represent the antidotes  (b) Red links show all the interference  links(complements of the antidote links)}\label{fig:topIAic}
%\end{center}
%%\vspace{-0.3cm}
%\end{figure}
The simplest example where non-Shannon inequalities are necessary, is shown in Fig. \ref{fig:full}. It is a multiple unicast setting with 11 messages, in which no receiver sees more than 2 interferers (only one-to-one alignment demands). Since the antidote graph in Fig. \ref{fig:full}(a) has too many edges, the interference graph, which is the complement of the antidote graph, and is therefore quite sparsely connected, is shown in Fig. \ref{fig:full}(b) instead. The interfering messages  at each receiver (equivalently, the antidotes missing at each receiver) are also listed in Fig. \ref{fig:full}(c) for convenience.

From an interference alignment perspective, the essence of the problem is captured by its alignment graph, shown in Fig. \ref{fig:full}(d) with solid black edges.  The alignment graph contains a node for each message, and each edge connects two interferers that are seen by the same destination, i.e., the pairs that are listed in Fig. \ref{fig:full}(c). Note the remarkable simplicity of the alignment graph, which involves only 7 alignment edges (solid black edges). As will be evident soon enough, the core of the problem, which makes non-Shannon inequalities necessary, has to do mainly with the four messages, $W_2, W_3, W_4, W_5$,  that comprise the inner diamond. Messages  $W_1, W_6$ create the requisite internal conflicts for the inner diamond to set off non-Shannon inequalities. The remaining messages, $W_7, W_8, W_9, W_{10}, W_{11}$, play only a marginal supporting role by demanding the alignments that bring the inner diamond into being.

\subsubsection{Necessity of Non-Shannon Inequalities}
The necessity of non-Shannon inequalities is established in the following theorem.

%We will use this instance to show the fact that there is a gap between the upper bound obtained by only Shannon-type information inequalities (also known as polymatroid upper bound) and the upper bound obtained by also utilizing the Zhang-Yeung non-Shannon-type information inequality \cite{ZY_Nonshannon}.

\begin{theorem}\label{theorem:nonShannon}
For the index coding instance of Fig. \ref{fig:full}, the best possible outer bound value on the symmetric capacity from only Shannon inequalities is $\frac{2}{5}$, whereas  the Zhang-Yeung non-Shannon-type information inequality yields the tighter outer bound value of $\frac{11}{28}$.
\end{theorem}

We present a sketch of the proof here. The details are in Section \ref{sec:nonShannon}.

From the alignment graph, we can see that the minimal internal conflict distance $\Delta = 2$. From Corollary 4.13 in \cite{Jafar_TIM}, we get an outer bound of $\frac{\Delta}{2\Delta + 1} = \frac{2}{5}$ which is based on polymatroid axioms. To  show that $\frac{2}{5}$ is the best outer bound that one can get with only  Shannon inequalities,  we use the bound in \cite{Arbabjolfaei_region}, which includes polymatroidal  (submodularity) axioms. As the bound is expressed in the form of a $T$ function defined over the power set of $\{1,2,\ldots,11\}$, we will find an explicit $T$ function with $R_i = \frac{2}{5}, i \in \{1,2,\ldots,11\}$ that satisfies all the polymatroidal axioms. Details are given in Section \ref{sec:nonShannon}.

We then show that the outer bound can be tightened to $\frac{11}{28} < \frac{2}{5}$. This proof consists of two parts. The first part uses the alignment chain $W_1 - W_3 - W_{4,5}$ (symmetrically $W_6 - W_4 - W_{2,3}$) to obtain a lower bound on the dimensions occupied by the entropic space of the triangle comprised of $W_2,W_3,W_4$ ($W_3,W_4,W_5$). This part only involves applying submodularity, i.e., Shannon-type information inequalities. The second part deals with the diamond comprised of $W_2,W_3,W_4,W_5$. Here we  use the Zhang-Yeung non-Shannon-type information inequality to obtain an upper bound for the dimensions occupied by the entropic space of the two triangles comprised of $W_2,W_3,W_4$ and $W_3,W_4,W_5$. Combining these two pieces yields the desired outer bound. %Details appear in Section \ref{sec:nonShannon}.

\subsubsection{Vector Space Interpretation and Linear Capacity}
Let us explore from a vector space perspective why  we need a non-polymatroidal inequality. Why do polymatroidal axioms not allow an outer bound smaller than $\frac{2}{5}$? Let us associate with each message $W_i$, the vector space ${\bf V}_i$. We will denote the vector space spanned by the union and intersection of spans of ${\bf V}_i$ and ${\bf V}_j$ as $({\bf V}_i, {\bf V}_j)$ and $({\bf V}_i\cap{\bf V}_j)$, respectively. Suppose we assign (normalized) subspace dimensions as $\dim({\bf V}_2) = \dim({\bf V}_3) = \dim({\bf V}_4) = \dim({\bf V}_5) = \frac{2}{5}$, $\dim({\bf V}_2, {\bf V}_3) =  \dim({\bf V}_2, {\bf V}_4) = \dim({\bf V}_3, {\bf V}_4) = \dim({\bf V}_3, {\bf V}_5) = \dim({\bf V}_4, {\bf V}_5) =  \frac{3}{5}$, and $\dim({\bf V}_2, {\bf V}_5) = \dim({\bf V}_2) + \dim({\bf V}_5) = \frac{4}{5}$ (because $W_2$ and $W_5$ conflict with each other, i.e., they cannot align) and finally, for the two triangles that create the diamond shape in the alignment graph, $\dim({\bf V}_2, {\bf V}_3, {\bf V}_4) = \dim({\bf V}_3, {\bf V}_4,{\bf V}_5) = \frac{4}{5}$. It is easy to verify that this dimension allocation satisfies all submodularity constraints, so an outer bound smaller than $\frac{2}{5}$ is not possible through polymatroidal axioms (submodularity) alone. Now let us see  why the submodularity bound must be loose, i.e., why the given dimension allocation cannot be satisfied by any vector space assignment.  Because $({\bf V}_3, {\bf V}_4)$  occupies $\frac{3}{5}$ dimensions, for the vector space $({\bf V}_2, {\bf V}_3, {\bf V}_4)$ to occupy $\frac{4}{5}$ dimensions, ${\bf V}_2$ must have $\frac{1}{5}$ new dimensions that have no intersection with $({\bf V}_3, {\bf V}_4)$. So it has only $\frac{1}{5}$ remaining dimensions that can intersect with $({\bf V}_3, {\bf V}_4)$. But it needs to intersect with each of ${\bf V}_3$ and ${\bf V}_4$ individually in  $\frac{2}{5}+\frac{2}{5}-\frac{3}{5}=\frac{1}{5}$ dimensions. Therefore ${\bf V}_2$ must intersect with ${\bf V}_3$ in the same $\frac{1}{5}$ dimensional space within which it intersects with ${\bf V}_4$. Therefore, the intersecting space of ${\bf V}_2$ with $({\bf V}_3, {\bf V}_4)$ must be the same as the intersecting space of ${\bf V}_3$ with ${\bf V}_4$, i.e., $({\bf V}_2\cap({\bf V}_3,{\bf V}_4))=({\bf V}_2 \cap {\bf V}_3) = ({\bf V}_2 \cap {\bf V}_4) = ({\bf V}_3 \cap {\bf V}_4)$. The same arguments can be made for the other triangle as well, i.e., $({\bf V}_5\cap({\bf V}_3,{\bf V}_4))=({\bf V}_5 \cap {\bf V}_3) = ({\bf V}_5 \cap {\bf V}_4) = ({\bf V}_3 \cap {\bf V}_4)$. However, this means that ${\bf V}_2$ and ${\bf V}_5$ must intersect in these dimensions as well. But this is a contradiction because $W_2$ conflicts with $W_5$.

\begin{figure}[h]
\begin{center}
\includegraphics[width= 3 in]{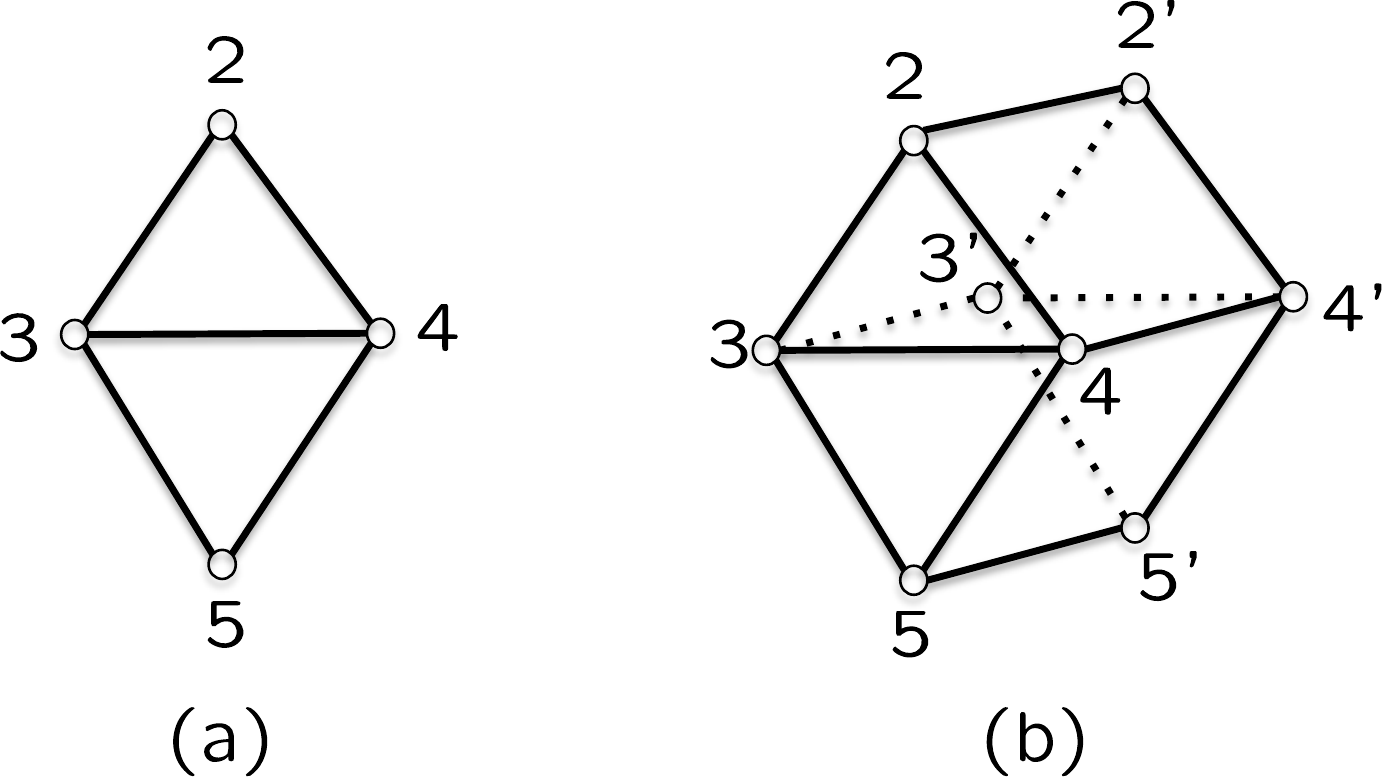}
\caption{ \it \small (a) Diamond part of the alignment graph and (b) Vamos matroid} \label{fig:vamos}
\end{center}
%\vspace{-0.3cm}
\end{figure}

As mentioned earlier, this contradiction cannot be captured by polymatroidal inequalities alone. However, the contradiction can be obtained as follows.
\begin{eqnarray*}
\mbox{dim}({\bf V}_2\cap {\bf V}_5)&\geq&\mbox{dim}({\bf V}_2\cap({\bf V}_3 \cap{\bf V}_4))+\mbox{dim}({\bf V}_5\cap({\bf V}_3 \cap{\bf V}_4))-\mbox{dim}({\bf V}_3\cap {\bf V}_4)\\
 &\geq& \mbox{dim}({\bf V}_2\cap{\bf V}_3)+\mbox{dim}({\bf V}_2\cap{\bf V}_4)-\mbox{dim}({\bf V}_2\cap({\bf V}_3, {\bf V}_4))\\
&&+\mbox{dim}({\bf V}_5\cap{\bf V}_3)+\mbox{dim}({\bf V}_5\cap{\bf V}_4)-\mbox{dim}({\bf V}_5\cap({\bf V}_3, {\bf V}_4))-\mbox{dim}({\bf V}_3,{\bf V}_4)
\end{eqnarray*}
\begin{eqnarray*}
\Rightarrow \mbox{dim}({\bf V}_2)+\mbox{dim}({\bf V}_5)-\mbox{dim}({\bf V}_2, {\bf V}_5)&\geq &\mbox{dim}({\bf V}_2)+\mbox{dim}({\bf V}_3)-\mbox{dim}({\bf V}_2, {\bf V}_3)\nonumber\\
&&+\mbox{dim}({\bf V}_2)+\mbox{dim}({\bf V}_4)-\mbox{dim}({\bf V}_2, {\bf V}_4)\nonumber\\
&&-\mbox{dim}({\bf V}_2)-\mbox{dim}({\bf V}_3, {\bf V}_4)+\mbox{dim}({\bf V}_2, {\bf V}_3, {\bf V}_4)\nonumber\\
&&\mbox{dim}({\bf V}_5)+\mbox{dim}({\bf V}_3)-\mbox{dim}({\bf V}_5, {\bf V}_3)\nonumber\\
&&+\mbox{dim}({\bf V}_5)+\mbox{dim}({\bf V}_4)-\mbox{dim}({\bf V}_5, {\bf V}_4)\nonumber\\
&&-\mbox{dim}({\bf V}_5)-\mbox{dim}({\bf V}_3, {\bf V}_4)+\mbox{dim}({\bf V}_3, {\bf V}_4, {\bf V}_5)\nonumber\\
&&-\mbox{dim}({\bf V}_3, {\bf V}_4)\nonumber
\end{eqnarray*}
\begin{eqnarray}
\Rightarrow&&\mbox{dim}({\bf V}_2,{\bf V}_3)+\mbox{dim}({\bf V}_2,{\bf V}_4)+\mbox{dim}({\bf V}_3,{\bf V}_4)+\mbox{dim}({\bf V}_3,{\bf V}_5)+\mbox{dim}({\bf V}_4,{\bf V}_5)\nonumber\\
&&~~~~~\geq\mbox{dim}({\bf V}_3)+\mbox{dim}({\bf V}_4)+
\mbox{dim}({\bf V}_2, {\bf V}_5)+\mbox{dim}({\bf V}_2, {\bf V}_3, {\bf V}_4)+\mbox{dim}({\bf V}_3, {\bf V}_4, {\bf V}_5)\label{eq:ingleton}
\end{eqnarray}
Note that the five terms on the left hand side correspond to the five edges of the diamond in Fig. \ref{fig:vamos}(a). Every step in this derivation is generally applicable to arbitrary vector subspaces ${\bf V}_2, {\bf V}_3, {\bf V}_4, {\bf V}_5$. In fact, what we have derived in (\ref{eq:ingleton}) is precisely the Ingleton inequality, which must be satisfied by any four vector subspaces. Plugging in the given dimension allocations we have on the left hand side a value of $\frac{3}{5}\times 5 = 3$ and on the right hand side a value of $\frac{2}{5}+\frac{2}{5}+\frac{4}{5}+\frac{4}{5}+\frac{4}{5}=\frac{16}{5}>3$, which violates (\ref{eq:ingleton}), thus producing a contradiction.

While we arrived at this example from an interference alignment perspective,  there are curious parallels to the Vamos matroid, previously used to establish the necessity of non-Shannon inequalities in the general network coding problem \cite{DFZ_Nonshannon} and in the multiple groupcast index coding problem \cite{Blasiak_Kleinberg_Lubetzky_2010}. Even the core of the alignment graph bears a resemblance to the Vamos matroid, as  illustrated in  Fig. \ref{fig:vamos}.

The Vamos matroid is an eight element ($\{2,2',3,3',4,4',5,5'\}$) matroid with rank 4. All 4-elements subsets are independent (naturally with rank 4) except the five planes $\{2,2',3,3'\}$,$\{2,2',4,4'\}$, \\ $\{3,3',4,4'\}$,$\{3,3',5,5'\}$,$\{4,4',5,5'\}$, which have rank 3 each. Now if we establish a correspondence ${\bf V}_2 \equiv \{2,2'\}, {\bf V}_3 \equiv \{3,3'\}, {\bf V}_4 \equiv \{4,4,'\}$ and ${\bf V}_5 \equiv \{5,5'\}$ and a normalization so that  rank 1 in the matroid is mapped to $\frac{1}{5}$ vector space dimensions, we find the dimension allocation that assigns $\frac{2}{5}$ dimensions per message for the diamond alignment graph matches  the rank function of the Vamos matroid. Now,  it is well known that the Vamos matroid is not representable (realizable by vector spaces) \cite{Oxley} and also not representable by entropic spaces, making this a natural example to require the use of the Ingleton  and non-Shannon inequalities. So, whether by coincidence or as a manifestation of a deeper mathematical property,  in our simplest example motivated by interference alignment,  the inner core of the alignment graph appears to be capturing the core dependence relationships of the Vamos matroid, with only marginal support from other parts of the alignment graph, instead of relying on all the circuits of the matroid. As a result, while the groupcast example based on the Vamos matroid used in \cite{Blasiak_Kleinberg_Lubetzky_2010} consists of  200 receiver nodes (the number of nodes associated with distinct circuits of the Vamos matroid), our unicast example involves only 11 receivers. On the other hand, because of the need for supporting messages our example does involve more messages (11 instead of 8). This is also because we want a unicast setting. As we will show later in this work, if we relax the problem to groupcast settings we can reduce the number of messages even further, to 6.

The vector space bounds based on the Ingleton inequality lead us to the linear capacity of our simplest example network.

\begin{theorem}\label{theorem:linear}
The symmetric linear capacity for the index coding instance shown in Fig. \ref{fig:full} is $\frac{5}{13}$.
\end{theorem}

\begin{figure}[h]
\begin{center}
\includegraphics[width= 5 in]{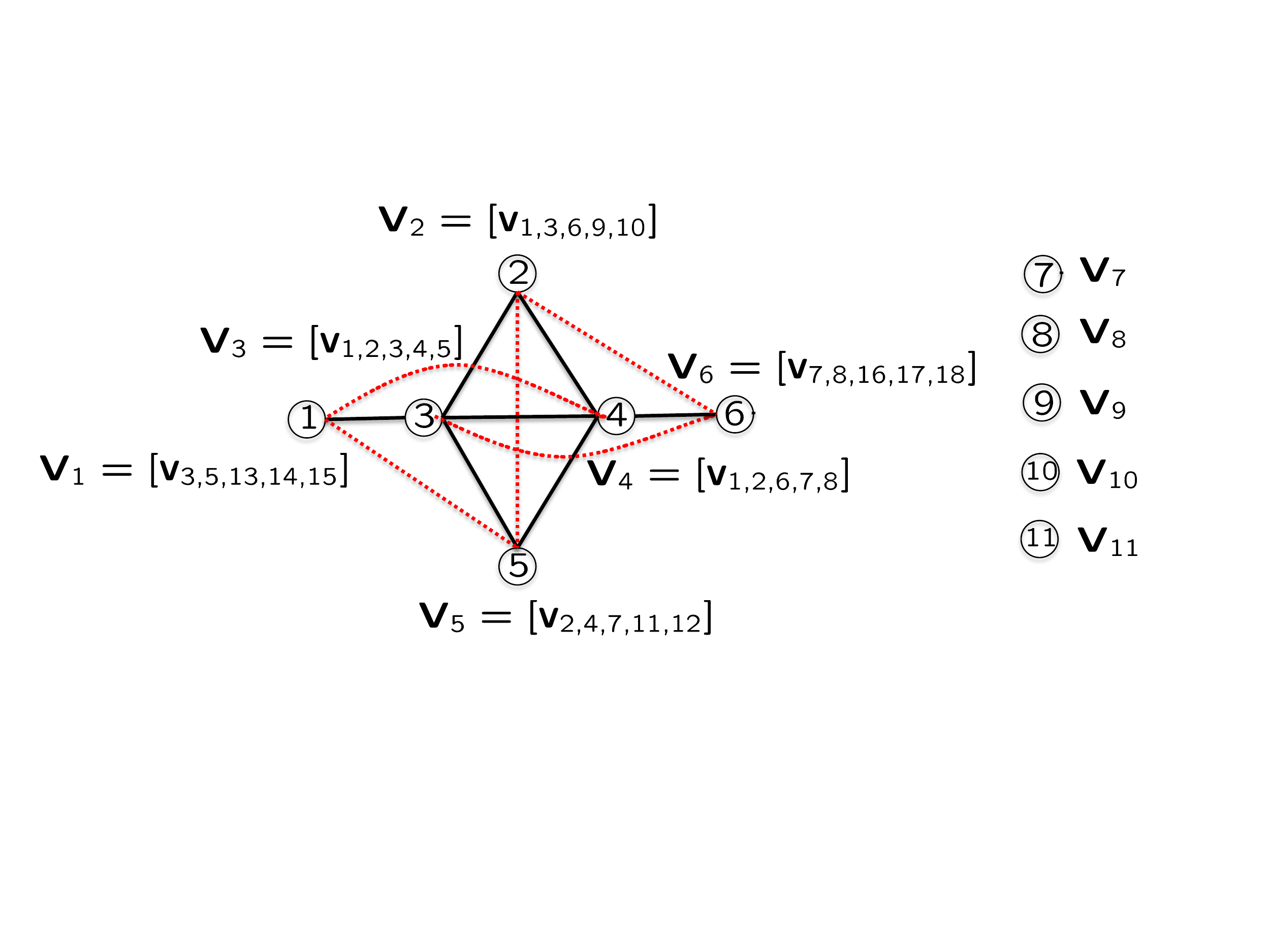}
\caption{ \it  \small A vector space assignment that avoids all conflicts. Symmetric rate
achieved is $5/13$ per message. ${\bf v}_i, {\bf v}_j$ is abbreviated as ${\bf v}_{i,j}$, etc.} \label{fig:ach}
\end{center}
%\vspace{-0.3cm}
\end{figure}

The outer bound follows as discussed previously, by the use of the Ingleton inequality which must be satisfied by all vector spaces. The proof is presented in Section \ref{sec:linear}. Here we give the achievable scheme. The achievable scheme is based on one-to-one alignment (see Fig. \ref{fig:ach}). The goal is to operate over $13$ channel uses and choose $5$ precoding vectors for each message, along which $5$ symbols for that message will be sent. The design uses insights from the outer bound to decide how much alignment should occur between the signal spaces. Here we mention some of the key values to facilitate the understanding of alignment. Denote the precoding matrix formed by $5$ precoding vectors for message $W_i$ as ${\bf V}_i$.  Then we  have, dim$({\bf V}_{i}, {\bf V}_j) = \frac{8}{13}$ whenever there is a solid black edge between $i,j$ in the alignment graph, and dim$({\bf V}_{2} \cap {\bf V}_{3} \cap {\bf V}_{4})$ = dim$({\bf V}_{3} \cap {\bf V}_{4} \cap {\bf V}_{5}) = \frac{1}{13}$. This is accomplished as follows. Generate $18+5\times 5=43$  vectors, each $13\times 1$, that are in general position (any 13 of them are linearly independent), over a sufficiently large field. For ${\bf V}_i, i \in \{1,2,\ldots, 6\}$, assign the first 18 vectors  according to Fig. \ref{fig:ach}. For ${\bf V}_i, i \in \{7,8,\ldots, 11\}$, assign $5$ of the remaining vectors to each. It is easy to check all internal conflicts are avoided and the space occupied by each alignment edge is $\frac{8}{13}$, leaving enough space, $1-\frac{8}{13}=\frac{5}{13}$, for the desired signal.

\subsubsection{There is no Simpler Example}
We now prove that this is indeed the simplest example where non-Shannon information inequalities are necessary. We will prove that for all multiple unicast index coding problems where each receiver is interfered by at most two messages (demanding only one-to-one alignments) and where the number of edges in the alignment graph is fewer than 7, the symmetric capacity is given by the internal conflict bound, $\frac{\Delta}{2\Delta+1}$, so that only Shannon-inequalities suffice. Half-rate-feasible networks are already solved in \cite{Jafar_TIM, Blasiak_Kleinberg_Lubetzky_2010} through only Shannon inequalities, so we will concern ourselves with only half-rate-infeasible settings in the following theorem.

\begin{theorem}\label{theorem:upto6}
For the class of half-rate-infeasible multiple-unicast index coding problems where each destination is interfered by at most two messages, if each alignment set contains fewer than or equal to 6 alignment edges, then the symmetric capacity is $\frac{\Delta}{2\Delta+1}$, where $\Delta$ is the minimum internal conflict distance.
\end{theorem}
\begin{figure}[h]
\begin{center}
\includegraphics[width= 6 in]{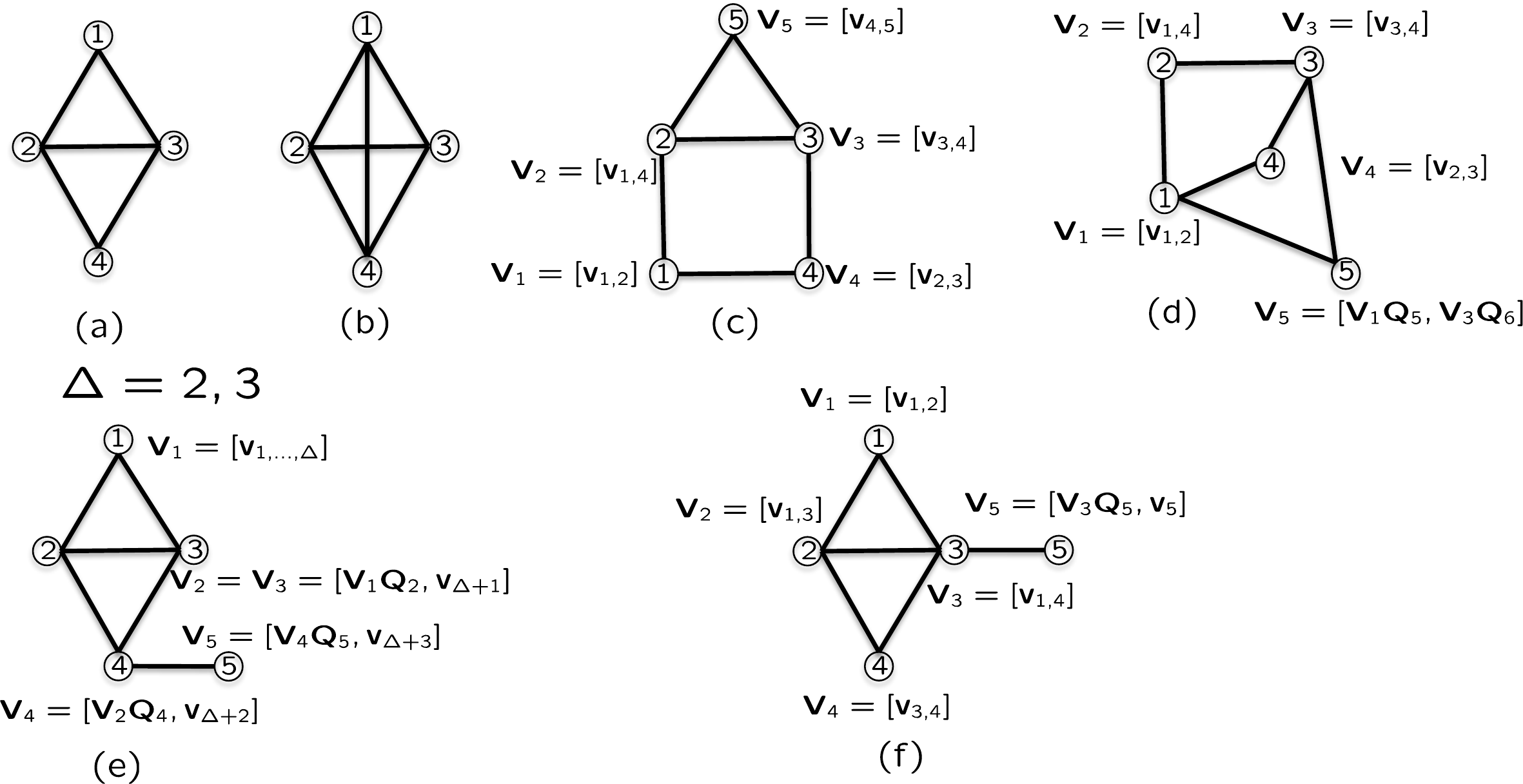}
\caption{ \it  \small All alignment graphs with 6 or fewer edges and containing  overlapping cycles. Also shown is a vector space assignment solution that avoids all conflicts at distance $\Delta$ or more.} \label{fig:upto6}
\end{center}
%\vspace{-0.3cm}
\end{figure}
{\textit{Proof:}} The case where each alignment set contains no overlapping cycles, i.e., no edge participates in more than one cycle,  is dealt with more generally (no constraints on the number of edges) in Section \ref{sec:cyc} in Theorem \ref{theorem:cyc}. Here we only consider the remaining cases where the alignment graph has overlapping cycles, as shown in Fig. \ref{fig:upto6}. We rely on a linear scheme over $2\Delta+1$ channel uses and send $\Delta$ symbols for each message. When $\Delta=1$, $\frac{\Delta}{2\Delta+1} = \frac{1}{3}$ can be achieved easily by multicast (CDMA) as each receiver is interfered by at most two messages. This is because over three channel uses, each receiver sees three generic linear equations in the three symbols (one desired, two interfering) that it is able to hear, from which it can resolve all three. Now consider $\Delta \geq 2$. The achievable scheme for each alignment set is shown in Fig. \ref{fig:upto6}. Note that case (a) is a subcase of case (e), allowing the same solution, and case (b) cannot have conflict distance more than 1. For all remaining cases, only (e) can have conflict distance 3. The ${\bf v}_i$ are $(2\Delta + 1)\times 1$  column vectors that are in general position, and all ${\bf Q}_i$ are generic transformations with dimension $\Delta \times (\Delta - 1)$. It is easily verified that each alignment edge occupies no more than $\Delta + 1$ dimensions and all internal conflicts are avoided.

Note that for case (f), we assume that if $W_5$ is interfered by two messages within the alignment set, the interfering messages are $W_1, W_2$ instead of $W_2,W_4$, without loss of generality. As this is a multiple unicast setting, it is not possible for $W_1,W_2$ and $W_2,W_4$ to be interference at $W_5$ simultaneously. But if we go to multiple groupcast setting, this is possible and we can construct an instance that exhibits similar properties as the instance shown in Fig. \ref{fig:full}. Specifically, we will present a simplest example where non-Shannon inequalities are needed for multiple \emph{groupcast} index coding problems in the next section.

\subsubsection{Multiple Groupcast: The Simplest Example}
\begin{figure}[h]
\begin{center}
\includegraphics[width=6.5 in]{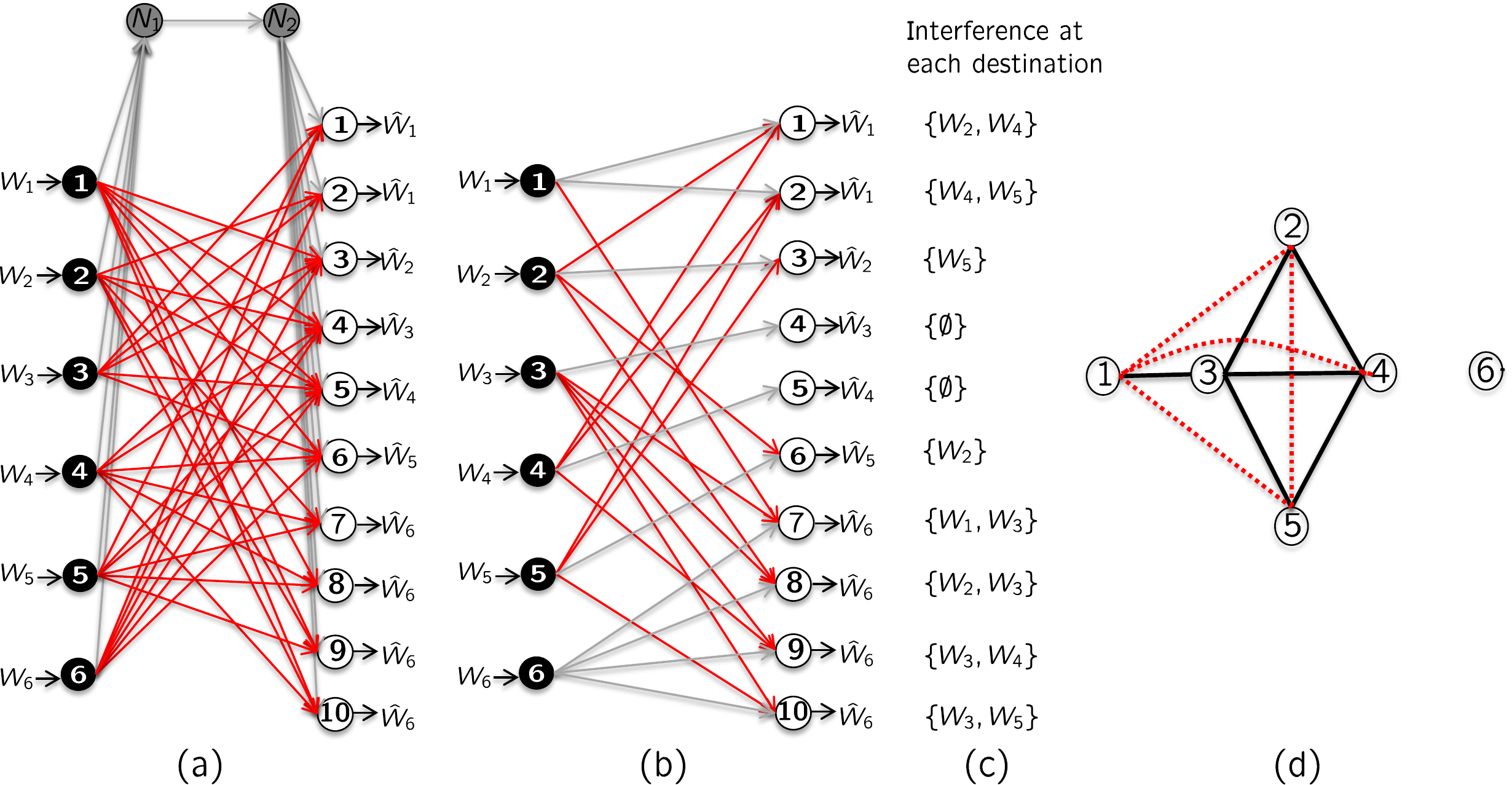}
\caption{ \it \small The Simplest Example of a Multiple Groupcast Index Coding Problem where Non-Shannon Inequalities are Necessary.  (a) Antidote Graph.  (b) Interference Graph. (c) List of interferers at each destination. (d) Alignment and Conflict Graphs. }\label{fig:group}
\end{center}
%\vspace{-0.3cm}
\end{figure}

As mentioned previously, Blasiak et al. have presented the first example (and the only example previously identified) of a multiple groupcast index coding problem in \cite{Blasiak_Kleinberg_Lubetzky_2010} where non-Shannon inequalities are necessary. The example presented by Blasiak et al. is based directly on the Vamos matroid, so that it contains 8 messages and 200 receivers (one for every  element of every circuit). In this section we use the interference alignment perspective to identify the simplest such example for multiple groupcast index coding. Our new criteria for the simplest example are the following.

\bigskip
\noindent{\bf Criteria for the Simplest Example}
\begin{enumerate}
\item Requires non-Shannon inequalities.
\item Each receiver sees no more than 2 interferers (one-to-one alignment demands).
\item Has the minimum number of edges in its alignment graph (among all examples that satisfy the first two criteria). %(alternatively, the minimum number of messages).
\end{enumerate}
Note that the restriction to multiple unicast is removed.

As we will show, the simplest example, shown in Fig. \ref{fig:group}, has only 6 messages (as opposed to 8  messages in the groupcast example of \cite{Blasiak_Kleinberg_Lubetzky_2010} and 11 messages in the simplest unicast example), only 6 alignment edges in the alignment graph (as opposed to 7 in the simplest unicast example), and a total of only 10 receivers (as opposed to 200  receivers in the groupcast example of \cite{Blasiak_Kleinberg_Lubetzky_2010} and 11 receivers in the simplest unicast example).  As the core part  (the diamond $W_2,W_3,W_4,W_5$) for the necessity of non-Shannon inequalities remains the same, the intuition and proofs follow  previous discussions. We establish the necessity of non-Shannon inequalities and find the linear capacity with the following theorem.

\begin{theorem}\label{theorem:group}
For the index coding instance shown in Fig. \ref{fig:group}, the best possible outer bound value on the symmetric capacity from only Shannon inequalities is $\frac{2}{5}$, whereas  the Zhang-Yeung non-Shannon-type information inequality yields the tighter outer bound value of $\frac{11}{28}$. Moreover, the symmetric linear capacity is $\frac{5}{13}$.
\end{theorem}

\begin{figure}[h]
\begin{center}
\includegraphics[width= 5 in]{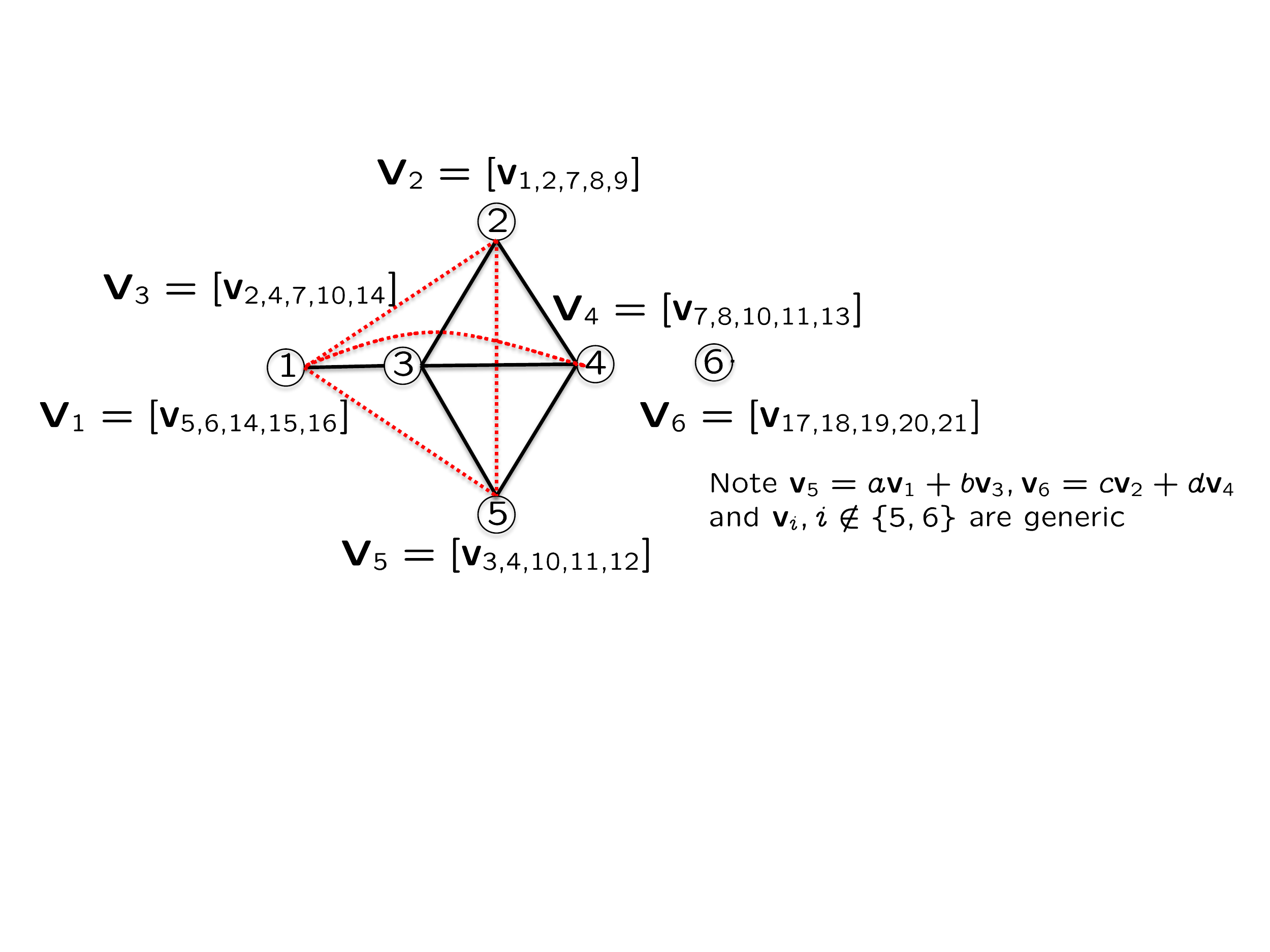}
\caption{ \it  \small A vector space assignment that avoids all conflicts. Symmetric rate
achieved is $5/13$ per message. ${\bf v}_i, {\bf v}_j$ is abbreviated as ${\bf v}_{i,j}$, etc. ${\bf v}_i, i \neq 5,6$ are generic $13 \times 1$ vectors and $a,b,c,d$ are generic scalars.} \label{fig:gach}
\end{center}
%\vspace{-0.3cm}
\end{figure}

The proof of Theorem \ref{theorem:group} follows along the same lines as the proofs for Theorem \ref{theorem:nonShannon} and Theorem \ref{theorem:linear}. Details are relegated to Section \ref{sec:group}. The optimal linear achievability scheme, shown in Fig. \ref{fig:gach}, has an  interesting aspect that even though the alignment demands are one-to-one, the optimal solution requires subspace alignments. Evidently, subspace alignment solutions may be required even when all demands are only one-to-one alignment demands.

%Rate $\frac{5}{13}$ is achieved by operating over $13$ channel uses and choosing $5$ precoding vectors for each message, along which $5$ symbols for that message will be sent. This is accomplished as shown in Fig. \ref{fig:gach}. It is easy to check all internal conflicts are avoided and the space occupied by each alignment edge is $\frac{8}{13}$, leaving enough space for the desired signal.

Next we prove that this is the simplest example for multiple-groupcast index coding problems. This is shown by the observation that all simpler cases are already solved without the need for non-Shannon inequalities. For half-rate-feasible networks, the capacity is already known and only Shannon inequalities are required \cite{Jafar_TIM}. For half-rate-infeasible networks, if the number of interferers seen by each destination is not more than 2, and there are fewer than 6 alignment edges in an alignment set, then the following theorem shows that the conflict bound (also based only on Shannon inequalities) is tight.

\begin{theorem}\label{theorem:groupupto5}
For the class of half-rate-infeasible multiple-groupcast index coding problems where each destination is interfered by at most two messages, if  each alignment set contains fewer than 6 alignment edges, then the symmetric capacity is $\frac{\Delta}{2\Delta+1}$, where $\Delta$ is the minimum internal conflict distance.
\end{theorem}

{\textit{Proof:}}  Among such cases, case (a) in Fig. \ref{fig:upto6} is the only one with overlapping cycles. The achievable scheme follows from case (e) with $\Delta = 2$ in Fig. \ref{fig:upto6}. Cases with no overlapping cycles are dealt with separately in Theorem \ref{theorem:cyc}.

{\it Remark:} It is not difficult to verify that the multiple groupcast index coding network of Fig. \ref{fig:group} is also the simplest network if instead of the minimum number of alignment edges, we require the minimum number of messages.

\subsection{Expanding the Interference Alignment Perspective}\label{sec:cyc}

Recall that \cite{Jafar_TIM} presents a characterization of the multiple groupcast index coding capacity for all instances where each connected component of the alignment graph, i.e., each alignment set, either does not contain a cycle or does not contain a fork. The next step to expand this class is  to also allow alignment sets to contain both cycles and forks. In our search for the simplest example, we were able to find such an extension. The result is presented as the next theorem.

\begin{theorem}\label{theorem:cyc}
For the class of half-rate-infeasible multiple-groupcast index coding problems where each destination is interfered by at most two messages, if each alignment set has no overlapping cycles, i.e., no two cycles share an edge, then the symmetric capacity is $\frac{\Delta}{2\Delta+1}$, where $\Delta$ is the minimum internal conflict distance.
\end{theorem}

In \cite{Jafar_TIM}, forks are handled by inheriting vectors from parent nodes, and cycles are handled by cyclic assignments of vector spaces.  Here, we  combine both techniques  to simultaneously deal with cycles and forks. %With this expansion of the interference alignment perspective, the next challenge will be to incorporate overlapping cycles on the one hand, and to deal with subspace alignments on the other. While the simple example constructed in this work sheds light on the challenges of overlapping cycles in alignment graphs,  the challenges of subspace alignments remain an intriguing direction for future work.

\begin{figure}[h]
\begin{center}
\includegraphics[width=5 in]{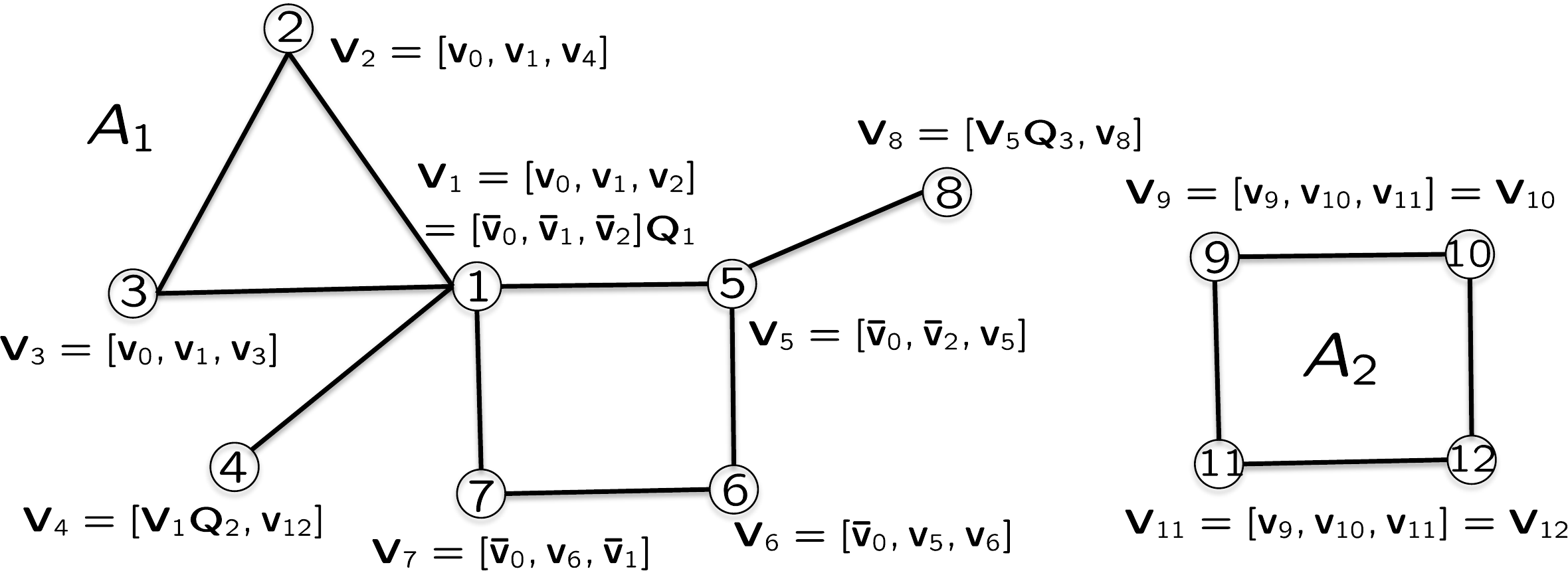}
\caption{\it \small An alignment graph showing a solution that avoids all conflicts at distance 3. Symmetric rate achieved is $3/7$ per message.}\label{fig:cycill}
\end{center}
%\vspace{-0.3cm}
\end{figure}

As an illustration of the result, an alignment graph is shown in Fig. \ref{fig:cycill}. While conflicts are not shown, the minimum internal conflict distance is assumed to be 3. Note that the assignment of precoding vectors avoids all conflicts at distance 3 or more. Two alignment sets are shown, labeled as $A_1, A_2$. $A_2$ cannot have an internal conflict because the minimum conflict distance is 3. All ${\bf v}_i$ are  $7 \times 1$ vectors in general position. ${\bf Q}_1$ is a generic $3 \times 3$ matrix and ${\bf Q}_2, {\bf Q}_3$ are  generic $3 \times 2$ matrices, meant to extract a generic 2 dimensional subspace from the connected node's signal space.  Each message occupies $3/7$ dimensions and any two adjacent messages occupy no more than $4/7$ dimensions, leaving the remaining $3/7$ dimensions for the desired signal. Note that for cycles, some randomly permuted common vectors are used combined with cyclicly assigned vectors. Detailed proof is presented in Section \ref{sec:cyct}.

\section{Discussion}
\begin{figure}[h]
\begin{center}
\includegraphics[width=5 in]{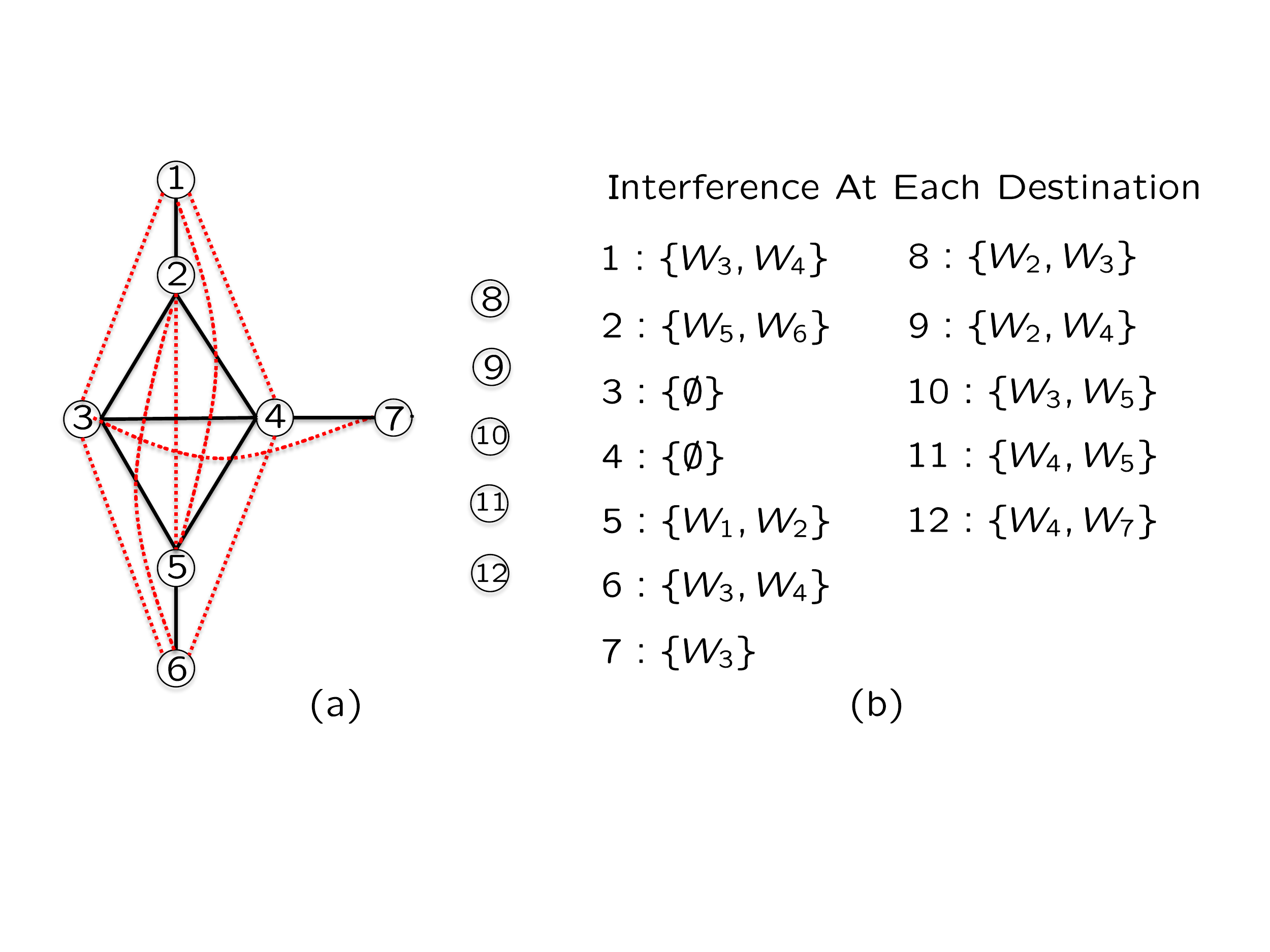}
\caption{\it \small A multiple unicast index coding problem with 8 edges in the alignment graph where non-Shannon inequalities are necessary. (a) Alignment and conflict graphs, and (b) Only the list of interferers at each destination is shown for simplicity.}\label{fig:exfull}
\end{center}
%\vspace{-0.3cm}
\end{figure}

While in this work we focused on the simplest instances, the necessity of non-Shannon inequalities can appear in different settings beyond the simplest examples. For example, Fig. \ref{fig:exfull} shows an interesting example with 8 alignment edges in the alignment graph. This is a multiple unicast index coding problem with 12 messages. While the inner diamond remains the same as  previous examples, the supporting edges appear in different positions relative to the diamond. Using the same approach, we  get corresponding results for this network --- although the best outer bound based on only Shannon inequalities is $\frac{2}{5}$, we can tighten it in this case to $\frac{13}{33}$ with the Zhang-Yeung non-Shannon type information inequality and the linear capacity, in this case, is $\frac{7}{18}$, which is shown using the  Ingleton inequality. The proof follows along similar lines of the proofs already presented.

While non-Shannon inequalities have been extensively studied, they are still not well understood both with regards to their fundamental character and their practical implications. The natural perspective provided by interference alignment can be quite valuable in both regards. It is tempting to interpret the non-Shannon inequalities, from an interference alignment perspective, as representing precisely the possibilities of subspace alignments. As the number of messages grows, so does the number of possibilities in how these messages can align with each other.  Much like the curious resemblance between the alignment graph of our  example which  is motivated by the interference alignment perspective, and the Vamos matroid,  there are noticeable parallels between the observations made through interference alignment and the fundamental information inequalities. As another example, consider  first the observation that while there are only finitely many information inequalities for upto 3 random variables, the number of information inequalities grows to infinity when 4 or more variables are involved.  This observation could be related to the existence of non-asymptotic interference alignment solutions for generic channels in the $K$ user interference channel for up to 3 users, but not beyond 3 users.

It is notable that in the degrees of freedom studies of wireless networks with generic channels, sufficient diversity, and full channel knowledge, the need for non-Shannon inequalities has not yet surfaced. The essential connection between the topological interference management problem and the index coding problem that is highlighted in \cite{Jafar_TIM}, does  link the results of this work to wireless networks with channel uncertainty. However, it is still not clear that non-Shannon inequalities are necessary for the topological interference management problem. This is because, unlike the index coding problem where all messages can be encoded together,  there is an additional constraint that the encoding of messages must be done in a distributed fashion in an interference network. It is possible that the additional constraint may manifest itself purely through submodularity to produce tighter outer bounds. The distributed coding requirement combined with the linear channel also favors the optimality of linear solutions in interference networks. Whether the linear coding capacity found for our simplest examples is the true capacity of these instance of the index coding problem, or for the corresponding instances of the topological interference management problem, remains a mystery.

\section{Proofs}
\subsection{Proof for Theorem \ref{theorem:nonShannon}}\label{sec:nonShannon}
The $T$ function is assigned as follows.
\begin{eqnarray}
&&T_\emptyset = 0, T_i = \frac{2}{5}, \forall i \in \{1,2,\ldots,11\} \\
&&T_{1,3} = T_{2,3} = T_{2,4} = T_{3,4} = T_{3,5} = T_{4,5} = T_{4,6} = \frac{3}{5}, T_{i,j} = \frac{4}{5} ~~\text{for all the other ${i,j}$ } \label{eqn:tff}\\
&&T_{1,2,3} = T_{1,3,5} = T_{1,3,4} = T_{2,3,4} = T_{3,4,5} = T_{2,3,5} = T_{2,4,5} = T_{3,4,6} = T_{2,4,6} = T_{4,5,6} = \frac{4}{5} \label{eqn:tf} \\
&&T_{i,j,k} = 1 ~~\text{for all the other $i,j,k$ } \\
&&T_{2,3,4,5} = \frac{4}{5}  ~~\text{and all unmentioned values of $T$ function are 1}.
\end{eqnarray}
It is trivial to check $R_i \leq T_{\{i\} \cup \mathcal{I}_i} - T_{\mathcal{I}_i}$ where $\mathcal{I}_i$ is the index set of the interfering messages at destination $i$. For $T$ function, it is easy to see that $T_\emptyset = 0, T_{1,2,\ldots,11} = 1$ and $T_{\mathcal{J}} \leq T_{\mathcal{K}}, \forall \mathcal{J} \subset \mathcal{K}$. We are only left to show the submodularity property, i.e., $T_{\mathcal{J}} + T_{\mathcal{K}} \geq T_{\mathcal{J} \cup {\mathcal{K}}} + T_{\mathcal{J} \cap \mathcal{K}},\forall \mathcal{J}, \mathcal{K}$. Without loss of generality, we assume $|\mathcal{J}| \leq |\mathcal{K}|$. We note that submodularity holds for two special cases trivially, i.e., $\mathcal{J} \cap \mathcal{K} = \emptyset$ or $\mathcal{J} \subset \mathcal{K}$. Henceforth we consider only the cases where $\mathcal{J}$ intersects with, but does not belong to $\mathcal{K}$, i.e., $|\mathcal{J}| \geq 2$.

If $|\mathcal{J}| = |\mathcal{K}| = 2$, $T_{\mathcal{J}} + T_{\mathcal{K}} \geq T_{\mathcal{J} \cup {\mathcal{K}}} + T_{\mathcal{J} \cap \mathcal{K}}$ holds when $T_{\mathcal{J}} = T_{\mathcal{K}} = \frac{3}{5}$ as $T_{\mathcal{J} \cup \mathcal{K}} = \frac{4}{5} ~~\text{from (\ref{eqn:tf})}$ and $T_{\mathcal{J} \cap \mathcal{K}} = T_{i} = \frac{2}{5}$. And also all the other cases as $T_{\mathcal{J}} + T_{\mathcal{K}} \geq \frac{3}{5} + \frac{4}{5} \geq 1 + \frac{2}{5}  \geq T_{\mathcal{J} \cup {\mathcal{K}}} + T_{\mathcal{J} \cap \mathcal{K}}$.

If $|\mathcal{J}| = 2, |\mathcal{K}| = 3$, submodularity also holds as $|\mathcal{J} \cap \mathcal{K}| = 1, T_{\mathcal{J} \cap \mathcal{K}} = \frac{2}{5}$ and $T_{\mathcal{J}} + T_{\mathcal{K}} \geq \frac{3}{5} + \frac{4}{5} \geq 1 + \frac{2}{5}  \geq T_{\mathcal{J} \cup {\mathcal{K}}} + T_{\mathcal{J} \cap \mathcal{K}}$. Similar reason applies when $|\mathcal{J}| = 2, |\mathcal{K}| \geq 4$.

If $|\mathcal{J}| = |\mathcal{K}| = 3$, submodularity holds whether $|\mathcal{J} \cap \mathcal{K}| = 1$ (trivial) or $|\mathcal{J} \cap \mathcal{K}| = 2$ as $T_{\mathcal{J} \cap \mathcal{K}} = \frac{2}{5}$ from (\ref{eqn:tff}) with the only exception $T_{2,3,5} + T_{2,4,5} = \frac{8}{5} \geq T_{2,3,4,5} + T_{2,5} = \frac{8}{5}$.

Consider $|\mathcal{J}| = 3, \mathcal{K} = \{2,3,4,5\}$. If further $T_{\mathcal{J}} = \frac{4}{5}$, all possible intersections belong to (\ref{eqn:tff}) and $T_{\mathcal{J} \cap \mathcal{K}} = \frac{3}{5}$, satisfying  submodularity. Otherwise $T_{\mathcal{J}} = 1$, $T_{\mathcal{J}} + T_{\mathcal{K}} \geq 1 + \frac{4}{5} \geq T_{\mathcal{J} \cup {\mathcal{K}}} + T_{\mathcal{J} \cap \mathcal{K}}$. This formula also holds for all the other cases where $|\mathcal{K}| \geq 4$.

For all  cases that remain, $T_\mathcal{J} = T_\mathcal{K} = 1$ and submodularity follows trivially. This completes the proof for polymadroid upper bound.

Now we proceed to the tighter outer bound where we will use non-Shannon information inequalities. The proof involves two parts, only the second part will involve a non-Shannon inequality. We start with the first part. Each solid black edge in the alignment graph represents an interference union at a desired destination. In order to leave $R$ dimensions for the desired message, the interference union must be compressed to a space with dimension smaller than $1-R$. This is made information theoretically rigorous as follows. Consider an alignment edge $(i,j)$ in the alignment graph and their interfering destination $k$ that desires $W_k$. From Fano's inequality, we have
\begin{eqnarray}
nR_k &\leq& I(W_k;\mathcal{S}^n, W_{i,j,k}^c) + o(n) \\
&\leq& I(W_k;\mathcal{S}^n | W_{i,j,k}^c) + o(n) \label{eqn:i1} \\
&\leq& H(\mathcal{S}^n | W_{i,j,k}^c) - H(\mathcal{S}^n | W_{i,j}^c) + o(n) \label{eqn:i}\\
&\leq& n - H(\mathcal{S}^n | W_{i,j}^c) + o(n) \label{eqn:i2} \\
\Rightarrow H(\mathcal{S}^n | W_{i,j}^c) &\leq& n(1-R_k) + o(n) \label{eqn:inf}
\end{eqnarray}
where (\ref{eqn:i1}) follows from the independence between the messages and (\ref{eqn:i2}) is due to the fact that $\mathcal{S}$ carries one symbol per channel use. We also get a byproduct from (\ref{eqn:i}),
\begin{eqnarray}
H(\mathcal{S}^n | W_{i,j,k}^c) \geq H(\mathcal{S}^n | W_{i,j}^c) + nR_k + o(n) \label{eqn:sep}
\end{eqnarray}
whose meaning is that the interference space $H(\mathcal{S}^n | W_{i,j}^c)$ is separable from the desired signal space.

Giving all the other messages except the desired $W_k$ as antidotes to destination $k$, we have
\begin{eqnarray}
nR_k &\leq& I(W_k;\mathcal{S}^n | W_{k}^c) + o(n)  \\
&\leq& H(\mathcal{S}^n | W_{k}^c) - H(\mathcal{S}^n | W_{\emptyset}^c) + o(n)\\
&\leq& H(\mathcal{S}^n | W_{k}^c) + o(n) \label{eqn:r}
\end{eqnarray}
where (\ref{eqn:r}) follows from the observation that $\mathcal{S}^n$ is a function of all the messages and this intuitively states that  the information contained in $\mathcal{S}^n$ cannot be smaller than the entropy of desired message $W_k$ when the uncertainty due to the other messages is not present.

For the alignment chain $W_1 - W_3 - W_{4,5}$, we have
\begin{eqnarray}
 H(\mathcal{S}^n | W_{1,3}^c) + H(\mathcal{S}^n | W_{3,4,5}^c) &\geq& H(\mathcal{S}^n | W_{3}^c) + H(\mathcal{S}^n | W_{1,3,4,5}^c) \label{eqn:t1} \\
&\geq& H(\mathcal{S}^n | W_{3}^c) + H(\mathcal{S}^n | W_{1,4,5}^c) \label{eqn:t2} \\
&\geq& H(\mathcal{S}^n | W_{3}^c) + nR_1 + H(\mathcal{S}^n | W_{4,5}^c)+o(n)  \label{eqn:t3} \\
&\geq& nR_3 +nR_1 + H(\mathcal{S}^n | W_{4,5}^c) + o(n) \label{eqn:t4}\\
\Rightarrow H(\mathcal{S}^n | W_{3,4,5}^c) &\geq& n(3R-1) + H(\mathcal{S}^n | W_{4,5}^c) + o(n)\label{eqn:t}
\end{eqnarray}
where (\ref{eqn:t1}) follows from submodularity, (\ref{eqn:t2}) is due to the fact that conditioning reduces entropy, (\ref{eqn:t3}) follows from (\ref{eqn:sep}) with $i=4,j=5,k=1$, (\ref{eqn:t4}) follows from (\ref{eqn:r}) with $k=3$ and (\ref{eqn:t}) is due to (\ref{eqn:inf}) with $i=1,j=3$ and because we are only interested in symmetric capacity.

Symmetrically, for the alignment chain $W_6 - W_4 - W_{2,3}$, we have
\begin{eqnarray}
H(\mathcal{S}^n | W_{2,3,4}^c) \geq n(3R-1) + H(\mathcal{S}^n | W_{2,3}^c)+o(n) \label{eqn:tt}
\end{eqnarray}
and adding (\ref{eqn:t}) and (\ref{eqn:tt}), we arrive at the final formula for the first part
\begin{eqnarray}
H(\mathcal{S}^n | W_{2,3,4}^c) + H(\mathcal{S}^n | W_{3,4,5}^c) \geq n(6R-2) + H(\mathcal{S}^n | W_{2,3}^c) + H(\mathcal{S}^n | W_{4,5}^c) +o(n) \label{eqn:1st}.
\end{eqnarray}
%Before going to the second part of the argument, we give a bound for $H(\mathcal{S}^n | W_{3,4}^c)$ that will be useful later. Consider the alignment chain $W_3 - W_4 - W_7$, we have
%\begin{eqnarray}
%H(\mathcal{S}^n | W_{3,4}^c) + H(\mathcal{S}^n | W_{4,7}^c) &\geq& H(\mathcal{S}^n | W_{4}^c) + H(\mathcal{S}^n | W_{3,4,7}^c) \label{eqn:d1} \\
%&\geq& H(\mathcal{S}^n | W_{4}^c) + H(\mathcal{S}^n | W_{3,7}^c) \label{eqn:d2} \\
%&\geq& H(\mathcal{S}^n | W_{4}^c) + nR_7 + H(\mathcal{S}^n | W_{3}^c)+o(n)  \label{eqn:d3} \\
%&\geq& nR_4 +nR_7 + nR_3  + o(n) \label{eqn:d4}\\
%\Rightarrow H(\mathcal{S}^n | W_{3,4}^c) &\geq& n(4R-1) + o(n)\label{eqn:d}
%\end{eqnarray}
%where (\ref{eqn:d1}) follows from submodularity, (\ref{eqn:d2}) is due to the fact that conditioning reduces entropy, (\ref{eqn:d3}) follows from (\ref{eqn:sep}) with $i=3,j=\emptyset,k=7$, (\ref{eqn:d4}) applies (\ref{eqn:r}) twice with $k=4,3$ and (\ref{eqn:d}) is due to (\ref{eqn:inf}) with $i=4,j=7$ and we are interested in symmetric capacity.

We proceed to the deal with the diamond. As mentioned before, we need the Zhang-Yeung non-Shannon-type information inequality for 4 random variables, stated as follows.
\begin{theorem}\label{theorem:zhang-yeung}
(Zhang-Yeung non-Shannon-type information inequality \cite{ZY_Nonshannon})
\begin{eqnarray}
3H(A,C) + 3H(A,D) + 3H(C,D) + H(B,C) + H(B,D) \notag \\
\geq 2H(C) + 2H(D) + H(A,B) + H(A) + H(B,C,D) + 4H(A,C,D)
\end{eqnarray}
\end{theorem}
We assign the random variables as follows,
\begin{eqnarray}
A = \mathcal{S}^n,W_{2,4,5}^c , B = \mathcal{S}^n,W_{2,3,5}^c , C = \mathcal{S}^n,W_{2,3,4}^c , D = \mathcal{S}^n,W_{3,4,5}^c.
\end{eqnarray}
Substituting  into the Zhang-Yeung non-Shannon-type information inequality, we have
\begin{eqnarray}
&&3H(\mathcal{S}^n,W_{2,4}^c) + 3H(\mathcal{S}^n,W_{4,5}^c) + 3H(\mathcal{S}^n,W_{3,4}^c) + H(\mathcal{S}^n,W_{2,3}^c) + H(\mathcal{S}^n,W_{3,5}^c) \notag \\
&&\geq 2H(\mathcal{S}^n,W_{2,3,4}^c) + 2H(\mathcal{S}^n,W_{3,4,5}^c) + H(\mathcal{S}^n,W_{2,5}^c) + H(\mathcal{S}^n,W_{2,4,5}^c) + H(\mathcal{S}^n,W_{3}^c) + 4H(\mathcal{S}^n,W_{4}^c) \\
\Rightarrow && 3H(\mathcal{S}^n | W_{2,4}^c) + 3H(\mathcal{S}^n | W_{4,5}^c) + 3H(\mathcal{S}^n | W_{3,4}^c) + H(\mathcal{S}^n | W_{2,3}^c) + H(\mathcal{S}^n | W_{3,5}^c) \notag \\
&& \geq 2H(\mathcal{S}^n | W_{2,3,4}^c) + 2H(\mathcal{S}^n | W_{3,4,5}^c) + H(\mathcal{S}^n | W_{2,5}^c) + H(\mathcal{S}^n | W_{2,4,5}^c) + H(\mathcal{S}^n | W_{3}^c) + 4H(\mathcal{S}^n | W_{4}^c) \label{eqn:n1}\\
\Rightarrow && 7n(1-R) + 3H(\mathcal{S}^n|W_{4,5}^c)+ H(\mathcal{S}^n | W_{2,3}^c) \geq 2H(\mathcal{S}^n | W_{2,3,4}^c) + 2H(\mathcal{S}^n | W_{3,4,5}^c) + 2H(\mathcal{S}^n | W_{2,5}^c) + 5nR +o(n) \label{eqn:n2}\nonumber\\
&&\\
\Rightarrow && 7n(1-R) + 3H(\mathcal{S}^n|W_{4,5}^c)+ H(\mathcal{S}^n | W_{2,3}^c) \geq 2H(\mathcal{S}^n | W_{2,3,4}^c) + 2H(\mathcal{S}^n | W_{3,4,5}^c) + 9nR + o(n)\label{eqn:n3}\\
\Rightarrow && n(7-16R) + 3H(\mathcal{S}^n|W_{4,5}^c)+ H(\mathcal{S}^n | W_{2,3}^c) \geq 2H(\mathcal{S}^n | W_{2,3,4}^c) + 2H(\mathcal{S}^n | W_{3,4,5}^c) +o(n)\label{eqn:2nd}
\end{eqnarray}
where (\ref{eqn:n1}) follows from the independence of the messages and that the number of messages is the same on the LHS and the RHS, the LHS of (\ref{eqn:n2}) follows from the fact that $(2,4),(3,4),(3,5)$ are all alignment edges and use (\ref{eqn:inf}), the RHS of (\ref{eqn:n2}) follows from the principle that conditioning reduces entropy and uses (\ref{eqn:r}) with $k=3,4$ and (\ref{eqn:n3}) follows from (\ref{eqn:sep}) with $i=2,j=\emptyset,k=5$ and from  (\ref{eqn:r}) with $k=2$.

Next, let us switch the value of $A,B$, i.e., $ A = \mathcal{S}^n,W_{2,3,5}^c , B = \mathcal{S}^n,W_{2,4,5}^c , C = \mathcal{S}^n,W_{2,3,4}^c , D = \mathcal{S}^n,W_{3,4,5}^c$ and similar to (\ref{eqn:2nd}), we have
\begin{eqnarray}
n(7-16R) + 3H(\mathcal{S}^n|W_{2,3}^c)+ H(\mathcal{S}^n | W_{4,5}^c) \geq 2H(\mathcal{S}^n | W_{2,3,4}^c) + 2H(\mathcal{S}^n | W_{3,4,5}^c) +o(n)\label{eqn:2nd2}.
\end{eqnarray}

\noindent Adding (\ref{eqn:2nd}) and (\ref{eqn:2nd2}) and dividing by 2 on both sides, we have the final formula for the second part,
\begin{eqnarray}
n(7-16R) + 2H(\mathcal{S}^n|W_{2,3}^c)+ 2H(\mathcal{S}^n | W_{4,5}^c) \geq 2H(\mathcal{S}^n | W_{2,3,4}^c) + 2H(\mathcal{S}^n | W_{3,4,5}^c) +o(n)\label{eqn:2ndf}.
\end{eqnarray}

\noindent Adding (\ref{eqn:2ndf}) with 2 times (\ref{eqn:1st}) and normalizing by $n$, we arrive at the conclusion
\begin{eqnarray}
7-16R \geq 2(6R-2) \Rightarrow R \leq \frac{11}{28}
\end{eqnarray}
which completes the proof.\hfill\QED
\subsection{Proof for Theorem \ref{theorem:linear}} \label{sec:linear}
The proof of the linear outer bound follows along similar lines as the information theoretical outer bound, and also consists of two parts. As linear capacity must be less than the actual capacity, the outer bound for the information theoretical outer bound can be directly applied here. We directly borrow the conclusion of the first part as follows according to (\ref{eqn:1st}),
\begin{eqnarray}
\dim({\bf V}_{2,3,4}) + \dim({\bf V}_{3,4,5}) &\geq& 6R -2 + \dim({\bf V}_{2,3}) + \dim({\bf V}_{4,5}) \label{eqn:l1}.
\end{eqnarray}
Note that the entropy space term $H(\mathcal{S}^n | W_{i,j,k}^c)$ is replaced by the vector space form ${\bf V}_{i,j,k}$.

For the second part, i.e., the diamond, we resort to the Ingleton inequality for a tighter linear bound. Note that there exist some joint distributions of four random variables that violate the Ingleton inequality \cite{HRSK_Inequ}, so the Ingleton bound does not hold information theoretically. However, it does hold for all vector spaces, so we can use it for this proof. From the Ingleton inequality (\ref{eq:ingleton}), we have
\begin{eqnarray}
&&\dim({\bf V}_{2,3}) + \dim({\bf V}_{2,4}) + \dim({\bf V}_{4,5}) + \dim({\bf V}_{3,5}) + \dim({\bf V}_{3,4})  \\
&& \geq \dim({\bf V}_{3}) + \dim({\bf V}_{4}) + \dim({\bf V}_{2,5}) + \dim({\bf V}_{2,3,4}) + \dim({\bf V}_{3,4,5}) \\
&\Rightarrow& \dim({\bf V}_{2,3}) +\dim({\bf V}_{4,5}) + 3(1-R)  \geq 2R + \dim({\bf V}_{2}) + \dim({\bf V}_{5}) + \dim({\bf V}_{2,3,4}) + \dim({\bf V}_{3,4,5}) \label{eqn:linf}\nonumber\\
&&\\
&\Rightarrow& \dim({\bf V}_{2,3}) +\dim({\bf V}_{4,5}) + 3 - 7R  \geq  \dim({\bf V}_{2,3,4}) + \dim({\bf V}_{3,4,5})\label{eqn:l3}
\end{eqnarray}
where (\ref{eqn:linf}) follows from (\ref{eqn:inf}) where $(2,4),(3,4),(3,5)$ are all interference edges, (\ref{eqn:r}) with $k=3,4$ and there is conflict between $W_2$ and $W_5$, (\ref{eqn:l3}) is due to (\ref{eqn:r}) with $k=2,5$.

Adding (\ref{eqn:l1}) and (\ref{eqn:l3}), we have
\begin{eqnarray}
3 - 7R \geq 6R -2  \Rightarrow R \leq \frac{5}{13}
\end{eqnarray}
which completes the proof. \hfill\QED
\subsection{Proof for Theorem \ref{theorem:group}}\label{sec:group}
The polymatroid upper bound $\frac{2}{5}$ is proved by explicitly assigning the spaces that correspond to each message and their unions with proper dimensions. Note that we need a groupcast version of $T$ function \cite{Arbabjolfaei_region}, which can be constructed similarly and follows the  same rules as defined in \cite{Arbabjolfaei_region}. The $T$ function is defined over the power set $\{1,2,3,4,5,6\}$ and we assign its values as follows. It satisfies the same constraints as specified in a previous  proof in Section \ref{sec:nonShannon}.
\begin{eqnarray}
&&T_\emptyset = 0, T_i = \frac{2}{5}, \forall i \in \{1,2,3,4,5,6\} \\
&&T_{1,3} = T_{2,3} = T_{2,4} = T_{3,4} = T_{3,5} = T_{4,5} = \frac{3}{5}, T_{i,j} = \frac{4}{5} ~~\text{for all the other ${i,j}$ }\\
&&T_{1,2,3} = T_{1,3,5} = T_{1,3,4} = T_{2,3,4} = T_{3,4,5} = T_{2,3,5} = T_{2,4,5} = \frac{4}{5}, T_{i,j,k} = 1 ~~\text{for all the other $i,j,k$ } \nonumber\\
&&\\
&&T_{2,3,4,5} = \frac{4}{5}  ~~\text{and all remaining unspecified values of $T$ function are set to 1}.
\end{eqnarray}

Next we proceed to the information theoretic outer bound. For alignment chain $W_1 - W_3 - W_{4,5}$, following similar steps as (\ref{eqn:t}), we have
\begin{eqnarray}
H(\mathcal{S}^n | W_{3,4,5}^c) &\geq& n(3R-1) + H(\mathcal{S}^n | W_{4,5}^c) + o(n)\label{eqn:gc1}.
\end{eqnarray}
Similarly, for alignment chain $W_1 - W_3 - W_{2,4}$,
\begin{eqnarray}
H(\mathcal{S}^n | W_{2,3,4}^c) &\geq& n(3R-1) + H(\mathcal{S}^n | W_{2,4}^c) + o(n)\label{eqn:gc2}.
\end{eqnarray}
For the diamond, we use Zhang-Yeung non-Shannon inequality to get the following (refer to (\ref{eqn:2nd})),
\begin{eqnarray}
2H(\mathcal{S}^n | W_{2,4}^c) + 2H(\mathcal{S}^n | W_{4,5}^c) + n(7-16R) \notag  \geq 2H(\mathcal{S}^n | W_{2,3,4}^c) + 2H(\mathcal{S}^n | W_{3,4,5}^c) +o(n) \label{eqn:gc3}.
\end{eqnarray}
Combining (\ref{eqn:gc1})(\ref{eqn:gc2})(\ref{eqn:gc3}) and normalizing by $n$, we have
\begin{eqnarray}
 7-16R \geq 12R - 4 \Rightarrow R \leq \frac{11}{28}.
\end{eqnarray}
Then we consider the linear capacity. With reference to (\ref{eqn:gc1}) and (\ref{eqn:gc2}), we have
\begin{eqnarray}
\dim({\bf V}_{2,3,4}) + \dim({\bf V}_{3,4,5}) &\geq& 6R -2 + \dim({\bf V}_{2,4}) + \dim({\bf V}_{4,5}). \label{eqn:g1}
\end{eqnarray}
With Ingleton inequality (similar to (\ref{eqn:l3})), we have
\begin{eqnarray}
\dim({\bf V}_{2,4}) +\dim({\bf V}_{4,5}) + 3 - 7R  \geq  \dim({\bf V}_{2,3,4}) + \dim({\bf V}_{3,4,5}). \label{eqn:g2}
\end{eqnarray}
Adding (\ref{eqn:g1})(\ref{eqn:g2}), we arrive at
\begin{eqnarray}
 3 - 7R  \geq 6R -2 \Rightarrow R \leq \frac{5}{13}.
\end{eqnarray}
\hfill\QED

\subsection{Proof for Theorem \ref{theorem:cyc}} \label{sec:cyct}
%{\it Remark:} Here non-overlapping cycle is defined as no edge (instead of vertex) is a part of two cycles in a graph.
%{\textit{Proof:}}
The outer bound is already available from Theorem 4.13 in \cite{Jafar_TIM}. We only prove the achievability of $\frac{\Delta}{2\Delta+1}$.

The goal is to operate over $2\Delta+1$ channel uses and choose $\Delta$ precoding vectors for each message, along which $\Delta$ symbols for that message will be sent.  A key idea here is that the precoding for each alignment set is designed independently. So we will describe the precoding vector design for each \emph{type} of alignment set. According to whether there are internal conflicts in each alignment set, we have two cases.

{\bf 1. Alignment sets with no internal conflicts:} \\
For each alignment set $A_i$ that has no internal conflicts, we randomly generate a $(2\Delta+1)\times \Delta$ matrix ${\bf V}(A_i)$.
\begin{eqnarray}
{\bf V}(A_i)&=&\mbox{rand}(2\Delta+1, \Delta)
\end{eqnarray}
where rand($a,b$) is a function that returns a randomly generated $a\times b$ matrix. The same precoding matrix ${\bf V}(A_i)$ will  be used by every message node in $A_i$.  That is, $\Delta$ symbols for each message $W\in A_i$ will be sent along the  $\Delta$ columns of ${\bf V}(A_i)$. %Next we describe precoder design for the  alignment sets that have internal conflicts. These are further classified as follows.
%(normalized, in the wireless case, to satisfy power constraints)

\bigskip
{\bf 2. Alignment sets with internal conflicts:} \\
\begin{enumerate}
\item From each alignment set $A_i$, arbitrarily choose one message node, say, $W_1(A_i)$ to be the center node. For each center node $W_1(A_i)$, randomly and independently generate a $(2\Delta+1)\times \Delta$ precoding matrix ${\bf V}_1(A_i)$  to be used by the node message.
\begin{eqnarray}
{\bf V}_1(A_i)&=&\mbox{rand}(2\Delta+1, \Delta),
\end{eqnarray}

\item Consider all the cycles incident on $W_1(A_i)$ and name them as $C_1$ to $C_K$ where $K$ is the total number of cycles incident on node $W_1(A_i)$. Enqueue all the nodes in each cycle. Let the length of cycle $k$ be  $|C_k|=l_k, k \in \{1,\ldots,K\}$. Label the messages in cycle $k$ (in order) as $W_1(C_k), W_2(C_k), \cdots, W_{l_k}(C_k)$. Randomly permutate the precoding matrix ${\bf V}_1(A_i)$,
\begin{eqnarray}
{\bf Q}_k(A_i) &=& \mbox{rand}(\Delta, \Delta)\\
{\bf V}^{k}_1(A_i) &=& {\bf V}_1(A_i) \mbox {\bf Q}_k(A_i) \\
  &=& [{\bf u}^k_1(C_k), {\bf u}^k_2(C_k), \ldots, {\bf u}^k_{{(\Delta - \lfloor l_k/2 \rfloor)}^+}(C_k), {\bf v}^k_1(C_k), {\bf v}^k_2(C_k), \ldots, {\bf v}^k_{ \Delta - {{(\Delta - \lfloor l_k/2 \rfloor)}^+} }(C_k)]\nonumber\\
  &&
\end{eqnarray}
Note that we divide the precoding vectors into two parts. The left ${(\Delta - \lfloor l_k/2 \rfloor)}^+$ vectors would be the common vectors assigned to all the nodes in cycle $k$ and right $\Delta - {{(\Delta - \lfloor l_k/2 \rfloor)}^+}$ vectors would be used in a cyclic fashion. If $\Delta \leq \lfloor l_k/2 \rfloor$, there would not exist common vectors and there exists internal conflicts within cycle $k$. Otherwise, there would not be internal conflicts within cycle $k$ and there are ${\Delta - \lfloor l_k/2 \rfloor}$ common precoding vectors shared by every node in cycle $k$. Also note that the column space is preserved by random permutation. Then randomly generate $l_k - \Delta + {{(\Delta - \lfloor l_k/2 \rfloor)}^+}$ vectors, each $(2\Delta+1)\times 1$, and call them ${\bf v}^k_{\Delta - {{(\Delta - \lfloor l_k/2 \rfloor)}^+}+1}(C_k), {\bf v}^k_{\Delta - {{(\Delta - \lfloor l_k/2 \rfloor)}^+}+2}(C_k), \cdots, {\bf v}^k_{l_k}(C_k)$. Now assign the vectors cyclically (subscripts modulo $l$) for cycle $k$ as follows:
%right vector number would be $\Delta$ and moreover,
%and there would be $\lfloor l_k/2 \rfloor$ right vectors
\begin{eqnarray}
W_1(C_k)&:& {\bf U}^k(C_k) , {\bf v}^k_1(C_k), {\bf v}^k_2(C_k), \cdots, {\bf v}^k_{\Delta - {{(\Delta - \lfloor l_k/2 \rfloor)}^+}}(C_k)\\
W_2(C_k)&:& {\bf U}^k(C_k) , {\bf v}^k_2(C_k), {\bf v}^k_3(C_k), \cdots, {\bf v}^k_{\Delta - {{(\Delta - \lfloor l_k/2 \rfloor)}^+}+1}(C_k)\\
W_3(C_k)&:& {\bf U}^k(C_k) , {\bf v}^k_3(C_k), {\bf v}^k_4(C_k), \cdots, {\bf v}^k_{\Delta - {{(\Delta - \lfloor l_k/2 \rfloor)}^+}+2}(C_k)\\
\vdots &:&\vdots\\
W_{l_k}(C_k)&:& {\bf U}^k(C_k) , {\bf v}^k_{l_k}(C_k), {\bf v}^k_1(C_k), \cdots, {\bf v}^k_{\Delta - {{(\Delta - \lfloor l_k/2 \rfloor)}^+}-1}(C_k),
\end{eqnarray}
where ${\bf U}^k(C_k) = [{\bf u}^k_1(C_k), {\bf u}^k_2(C_k), \ldots, {\bf u}^k_{{(\Delta - \lfloor l_k/2 \rfloor)}^+}(C_k)]$ represents all the common vectors.
Note that this construction ensures that any adjacent two nodes share ${{(\Delta - \lfloor l_k/2 \rfloor)}^+} + \Delta - {{(\Delta - \lfloor l_k/2 \rfloor)}^+} - 1 = \Delta - 1$ dimensional precoding space. And message nodes that are connected by a (minimum) path of two edges in the cycle  have an overlap of $\Delta-2$ dimensions, message nodes that are connected by a path of three edges  have an overlap of $\Delta-3$ dimensions, and so on, so that messages that are connected by a path of $\Delta$ edges (or more) have no overlap. Thus, all conflicts within the cycle (if exist) are avoided.

\item Now we proceed to all the other nodes that are connected to $W_1(A_i)$ but do not form a cycle with $W_1(A_i)$. Enqueue all these nodes.
For each such node $W_j(A_i), j\neq 1$, precoding matrix is generated as:
\begin{eqnarray}
{\bf Q}_j(A_i)&=&\mbox{rand}(\Delta, \Delta-1)\\
{\bf V}_j(A_i)&=&[{\bf V}_{1}(A_i){\bf Q}_j(A_i)~~~~ \mbox{rand}(2\Delta+1,1)]
\end{eqnarray}

The random matrix ${\bf Q}_j(A_i)$ is simply meant to choose a generic $\Delta-1$ dimensional subspace from ${\bf V}_{1}(A_i)$. This is appended with an independently generated vector that will  (with high probability over a sufficiently large field) be in general position and linear independent with respect to $ {\bf V}_{1}(A_i) {\bf Q}_j(A_i)$. Thus, ${\bf V}_{1}(A_i)$ and $ {\bf V}_{j}(A_i)$ are connected by an edge in the alignment graph and have a $\Delta-1$ dimensional overlap between their signal spaces.
%An example of the construction is provided in Fig. \ref{fig:tree}(c) where the alignment set has no cycles, $\Delta=2$ and message $W_5$ is chosen as the root node.

\item Operations related to $W_1(A_i)$ are done. Dequeue and consider the next message node $W_j(A_i)$ in the queue. Now $W_j(A_i)$ becomes the new center node and go back to step 2 to deal with all the unassigned nodes. When the queue is empty again, the precoding matrix assignment is completed. Since the alignment graph does not have overlapping cycles, messages that are connected by a path (may cross multiple cycles) of $\Delta$ edges (or more) have no overlap with such construction as common vectors within different cycles are in generic positions because of the random permutation and other new appended vectors are generic naturally.

\end{enumerate}
%(i.e., probability 1 in the wireless case, and a probability that can be as close to 1 as needed over a sufficiently large field in the wired case)
After completing the construction, we are left to show that with high probability  the desired signals at each destination have no overlap with the interference. Without loss of generality we will assume each destination desires one message.

Consider a destination whose interfering messages come from an alignment set that has no internal conflicts. Thus, all interfering messages span the same $\Delta$ dimensional space, and the desired signal (because it belongs to a different alignment set) spans an independently generated $\Delta$ dimensional space. Since the overall number of dimensions is $2\Delta+1$, with high probability these two spaces have no overlap.

Henceforth we consider only destinations whose interfering messages come from an alignment set that has  internal conflicts.

Suppose the desired message, say $W_i$, sees two interferers $W_j, W_k$. Then $W_j, W_k$ must be connected by an edge in the alignment graph. Therefore, they must have a $\Delta-1$ dimensional overlap whether they are a part of a cycle or not, so that together they must span $\Delta+\Delta-(\Delta-1)=\Delta+1$ dimensions. Further, if $W_i$ is in the same alignment set, then  $W_j, W_k$ must be at least $\Delta$ edges away from $W_i$, so that with high probability the union of the spans of ${\bf V}_j, {\bf V}_k$ is in general position with respect to ${\bf V}_i$. Since the total space is $2\Delta+1$ dimensional, it is big enough to accommodate the interference and the desired signal without forcing them to overlap. Thus, the desired signal does not overlap with interference with high probability.  If the message $W_i$ is in a different alignment set then again its signal space is independently generated and with high probability has no overlap with the space spanned by the interference. If the message $W_i$ sees only one interferer, $W_j$, then once again because $W_i, W_j$ are at least $\Delta$ edges apart (or belong to different alignment sets), the signal spaces ${\bf V}_i, {\bf V}_j$ have no overlap with high probability.\hfill\QED

\appendix
\section*{Appendix}
\section{Summary of Relevant Definitions}\label{sec:def}
The index coding problem consists of $S$ source nodes, labeled $S_j, j \in \{1,2,\ldots, S\}$, $D$ destination nodes, labeled $D_i, i\in \{1,2,\ldots,D\}$ and two additional nodes, labeled as $N_1, N_2$, that are connected by a unit capacity edge going from $N_1$ to $N_2$, known as the bottleneck link. There is an infinite capacity link from every source to the node $N_1$, and an infinite capacity link from $N_2$ to every destination node. What it means is simply that $N_1$ knows all the messages, so all the coding is performed at $N_1$, and the output of the bottleneck link is available to all destination nodes.

Source node $S_j$ has a set of independent messages, $\mathcal{W}({S_j})$, that it wants to send to their desired destinations. Destination node $D_i$ has a set of independent messages $\mathcal{W}(D_i)$ that it desires. We distinguish between multiple groupcast settings where each message can be desired by multiple destinations and multiple unicast settings where each message cannot be desired by more than one destination. The side information structure is defined by the antidote matrix $\mathcal{A} = [a_{ij}]_{D \times S}$ of zeros and ones where $a_{ij}=1$ means a direct link of infinite capacity exists from $S_j$ to $D_i$ and provides $\mathcal{W}(S_j)$ to $D_i$, otherwise $a_{ij}=0$ and no path exists from $S_j$ to $D_i$ except through the bottleneck link. To avoid degenerate cases, we assume desired message are not available as side information and must pass through the bottleneck link, i.e., $a_{ij} = 0$ whenever $\mathcal{W}(D_i)\cap\mathcal{W}(S_j)\neq\phi$. The bottleneck link is able to transmit one symbol from $\mathbb{GF}$ each channel use and the symbol transmitted is denoted as $\mathcal{S}$.

Coding schemes, probability of error, achievable rates and capacity region are defined in the standard information theoretic sense of vanishing probability of error. We are interested in the symmetric capacity $R$ normalized by the capacity of the bottleneck link, $\log(|\mathbb{GF}|)$. Although the choice of field is irrelevant to the normalized capacity of the index coding problem, we assume the field is large to simplify the design of achievable scheme. Throughout the paper, we will use the notion of linear schemes defined in detail in Appendix A of \cite{Jafar_TIM}. Linear capacity is defined similar to capacity, but with the constraint that coding schemes must be linear.

We proceed to the definition of alignment and conflict graphs for the index coding problem.
\begin{enumerate}
\item {\bf Alignment Graph:} Messages $W_i$ and $W_j$ are connected with a solid black edge if the source(s) of both these messages are not available as antidotes to a destination that desires message $W_k\notin\{W_i, W_j\}$. %In this work we will focus primarily on settings where there are at most two messages that a destination does not know as antidotes.
\item{\bf Conflict Graph:} Each message $W_i$ is connected by a dashed red edge to  all other messages whose sources are are not available as antidotes to a destination that desires message $W_i$.
\end{enumerate}
Also let us define {\bf Alignment Set} to be each connected component (through solid black edges) of an alignment graph and
{\bf Internal Conflict} which refers to the occasion where two messages that belong to the same alignment graph have a conflict (dashed red) edge between them. If message $W_j$ is not available as antidote to destination $i$, we also call $W_j$ interference for destination $i$. $W_i,W_j$ is abbreviated as $W_{i,j}$, etc. $W_{i,j}^c$ is used to denote all the messages except $W_{i,j}$.

Finally, we say that the alignment graph has non-overlapping cycles when no edge is a part of two cycles in the graph.

\bibliographystyle{IEEEtran}
\bibliography{Thesis}

% Generated by IEEEtran.bst, version: 1.13 (2008/09/30)
\begin{thebibliography}{10}
\providecommand{\url}[1]{#1}
\csname url@samestyle\endcsname
\providecommand{\newblock}{\relax}
\providecommand{\bibinfo}[2]{#2}
\providecommand{\BIBentrySTDinterwordspacing}{\spaceskip=0pt\relax}
\providecommand{\BIBentryALTinterwordstretchfactor}{4}
\providecommand{\BIBentryALTinterwordspacing}{\spaceskip=\fontdimen2\font plus
\BIBentryALTinterwordstretchfactor\fontdimen3\font minus
  \fontdimen4\font\relax}
\providecommand{\BIBforeignlanguage}[2]{{%
\expandafter\ifx\csname l@#1\endcsname\relax
\typeout{** WARNING: IEEEtran.bst: No hyphenation pattern has been}%
\typeout{** loaded for the language `#1'. Using the pattern for}%
\typeout{** the default language instead.}%
\else
\language=\csname l@#1\endcsname
\fi
#2}}
\providecommand{\BIBdecl}{\relax}
\BIBdecl

\bibitem{Birk_Kol}
Y.~Birk and T.~Kol, ``{Informed-source coding-on-demand ({ISCOD}) over
  broadcast channels},'' in \emph{Proceedings of the Seventeenth Annual Joint
  Conference of the IEEE Computer and Communications Societies, IEEE
  INFOCOM'98}, vol.~3, 1998, pp. 1257--1264.

\bibitem{Jafar_FnT}
S.~Jafar, ``Interference alignment: A new look at signal dimensions in a
  communication network,'' in \emph{Foundations and Trends in Communication and
  Information Theory}, 2011, pp. 1--136.

\bibitem{Jafar_TIM}
\BIBentryALTinterwordspacing
S.~A. Jafar, ``{Topological Interference Management through Index Coding},''
  \emph{ArXiv:1301.3106}, Jan. 2013. [Online]. Available:
  \url{http://arxiv.org/abs/1301.3106}
\BIBentrySTDinterwordspacing

\bibitem{Maleki_Cadambe_Jafar}
H.~Maleki, V.~Cadambe, and S.~Jafar, ``Index coding -- an interference
  alignment perspective,'' \emph{ISIT 2012, Preprint of Full Paper available at
  ArXiv:1205.1483}, 2012.

\bibitem{Rouayheb_Sprintson_Georghiades}
S.~Rouayheb, A.~Sprintson, and C.~Georghiades, ``{On the Index Coding Problem
  and Its Relation to Network Coding and Matroid Theory},'' \emph{IEEE
  Transactions on Information Theory}, vol.~56, no.~7, pp. 3187--3195, July
  2010.

\bibitem{Effros_Rouayheb_Langberg}
M.~{Effros}, S.~{El Rouayheb}, and M.~{Langberg}, ``{An Equivalence between
  Network Coding and Index Coding},'' \emph{ArXiv:1211.6660}, Nov. 2012.

\bibitem{Yossef_Birk_Jayram_Kol_Trans}
{Z. Bar-Yossef and Y. Birk and T. S. Jayram and T. Kol}, ``{Index Coding With
  Side Information},'' \emph{IEEE Trans. on Information Theory}, vol.~57,
  no.~3, pp. 1479 -- 1494, March 2011.

\bibitem{Blasiak_Kleinberg_Lubetzky_2010}
\BIBentryALTinterwordspacing
A.~Blasiak, R.~Kleinberg, and E.~Lubetzky, ``Index coding via linear
  programming,'' \emph{ArXiv:1004.1379}, April 2010. [Online]. Available:
  \url{http://www.cs.cornell.edu/$\sim$ablasiak/papers/bkl-beta.pdf}
\BIBentrySTDinterwordspacing

\bibitem{Tehrani_Dimakis_Neely}
A.~Tehrani, A.~Dimakis, and M.~Neely, ``Bipartite index coding,'' in
  \emph{Proceedings of International Symposium on Information Theory (ISIT)},
  2012.

\bibitem{Alex_local}
\BIBentryALTinterwordspacing
K.~Shanmugam, A.~Dimakis, and M.~Langberg, ``{Local Graph Coloring and Index
  Coding},'' \emph{ArXiv:1301.5359}, Jan. 2013. [Online]. Available:
  \url{http://arxiv.org/abs/1301.5359}
\BIBentrySTDinterwordspacing

\bibitem{Alon_Hasidim_Lubetzky_Stav_Weinstein}
\BIBentryALTinterwordspacing
N.~Alon, A.~Hasidim, E.~Lubetzky, U.~Stav, and A.~Weinstein, ``Broadcasting
  with side information,'' \emph{ArXiv:0806.3246}, Jun. 2008. [Online].
  Available: \url{http://arxiv.org/pdf/0806.3246v1.pdf}
\BIBentrySTDinterwordspacing

\bibitem{Lubetzky_Nonlinear}
E.~Lubetzky and U.~Stav, ``Non-linear index coding outperforming the linear
  optimum,'' \emph{IEEE Trans. Inf. Theory}, vol.~55, no.~8, pp. 3544 -- 3551,
  Aug. 2009.

\bibitem{Blasiak_Kleinberg_Lubetzky_2011}
\BIBentryALTinterwordspacing
A.~Blasiak, R.~Kleinberg, and E.~Lubetzky, ``Lexicographic products and the
  power of non-linear network coding,'' \emph{ArXiv:1108.2489}, Aug. 2011.
  [Online]. Available: \url{http://arxiv.org/abs/1108.2489}
\BIBentrySTDinterwordspacing

\bibitem{Arbabjolfaei_region}
\BIBentryALTinterwordspacing
F.~Arbabjolfaei, B.~Bandemer, Y.~Kim, E.~Sasoglu, and L.~Wang, ``{On the
  Capacity Region for Index Coding},'' \emph{ArXiv:1302.1601}, Feb. 2013.
  [Online]. Available: \url{http://arxiv.org/abs/1302.1601v1}
\BIBentrySTDinterwordspacing

\bibitem{Naderi_Avestimehr}
N.~Naderializadeh and A.~S. Avestimehr, ``Interference networks with no csit:
  Impact of topology,'' \emph{ArXiv}, vol. abs/1302.0296, 2013.

\bibitem{Y_ITNC}
R.~W. Yeung, \emph{Information Theory and Network Coding}.\hskip 1em plus 0.5em
  minus 0.4em\relax Springer, 2008.

\bibitem{HRSK_Inequ}
{D. Hammer and A. E. Romashchenko and A. Shen, and N. K. Vereshchagin},
  ``{Inequalities for Shannon entropy and Kolmogorov complexity},'' \emph{J.
  Comput. Syst. Sci.}, vol.~60, pp. 442--464, 2000.

\bibitem{ZY_Nonshannon}
Z.~Zhang and R.~W. Yeung, ``{On characterization of entropy function via
  information inequalities},'' \emph{IEEE Trans. Inf. Theory}, vol.~44, no.~4,
  pp. 1440 -- 1452, Jul. 1998.

\bibitem{Makarychev}
K.~Makarychev, Y.~Makarychev, A.~Romashchenko, and N.~Vereshchagin, ``{A new
  class of non-Shannon-type inequalities for entropies},'' \emph{Communications
  in Information and Systems}, vol.~2, no.~2, pp. 147 -- 166, December 2002.

\bibitem{Matus_Infty}
F.~Matus, ``Infinitely many information inequalities,'' in \emph{Proceedings of
  International Symposium on Information Theory (ISIT)}, 2007, pp. 41 -- 44.

\bibitem{DFZ_Nonshannon4}
\BIBentryALTinterwordspacing
R.~Dougherty, C.~Freiling, and K.~Zeger, ``{Non-Shannon information
  inequalities in four random variables},'' \emph{ArXiv:1104.3602}, April 2011.
  [Online]. Available: \url{http://arxiv.org/abs/1104.3602v1}
\BIBentrySTDinterwordspacing

\bibitem{Ingleton}
A.~W. Ingleton, ``{Representation of matroids in combinatorial mathematics and
  its applications},'' \emph{Combinatorial Mathematics and Its Applications},
  vol.~44, pp. 149 -- 167, Jul. 1971.

\bibitem{Kinser}
R.~Kinser, ``{New inequalities for subspace arrangements},'' \emph{J. Combin.
  Theory}, vol. 118, pp. 152 -- 161, Jan. 2011.

\bibitem{DFZ_Rank}
\BIBentryALTinterwordspacing
R.~Dougherty, C.~Freiling, and K.~Zeger, ``{Linear rank inequalities on five or
  more variables},'' \emph{ArXiv:0910.0284}, July 2010. [Online]. Available:
  \url{http://arxiv.org/abs/0910.0284v3}
\BIBentrySTDinterwordspacing

\bibitem{DFZ_Matroids}
------, ``{Network coding and matroid theory},'' \emph{Proc. IEEE}, vol.~99,
  no.~3, pp. 388 -- 405, Mar. 2011.

\bibitem{CG_Duality}
T.~H. Chan and A.~Grant, ``{Dualities between entropy functions and network
  codes},'' \emph{IEEE Trans. Inf. Theory}, vol.~54, no.~10, pp. 4470 -- 4487,
  Oct. 2008.

\bibitem{DFZ_Nonshannon}
R.~Dougherty, C.~Freiling, and K.~Zeger, ``{Networks, matroids, and non-Shannon
  information inequalities},'' \emph{IEEE Trans. Inf. Theory}, vol.~53, no.~6,
  pp. 1949 -- 1969, Jun. 2007.

\bibitem{Y_Entropy}
R.~W. Yeung, ``{Facets of Entropy},'' \emph{IEEE Information Theory Society
  Newsletter}, vol.~62, no.~8, pp. 6 -- 16, December 2012.

\bibitem{C_Entropy}
T.~H. Chan, ``{Recent Progresses in Characterising Information Inequalities},''
  \emph{Entropy}, vol.~13, pp. 379 -- 401, 2011.

\bibitem{Oxley}
J.~G. Oxley, \emph{Matroid Theory}.\hskip 1em plus 0.5em minus 0.4em\relax New
  York: Oxford Univ. Press, 1992.

\end{thebibliography}
\end{document}